\pdfoutput=1
\documentclass[sigconf]{acmart}
\usepackage{todonotes}
\usepackage{enumitem,varwidth, hyperref}

\newcommand{\change}[1]{{#1}}

\usepackage{array}
\newcolumntype{C}[1]{>{\centering\let\newline\\\arraybackslash\hspace{0pt}}m{#1}}

\AtBeginDocument{%
  \providecommand\BibTeX{{%
    \normalfont B\kern-0.5em{\scshape i\kern-0.25em b}\kern-0.8em\TeX}}}

\copyrightyear{2023} 
\acmYear{2023} 
\setcopyright{acmlicensed}
\acmConference[CHI '23]{Proceedings of the 2023 CHI Conference on Human Factors in Computing Systems}{April 23--28, 2023}{Hamburg, Germany}
\acmBooktitle{Proceedings of the 2023 CHI Conference on Human Factors in Computing Systems (CHI '23), April 23--28, 2023, Hamburg, Germany}
\acmPrice{15.00}
\acmDOI{10.1145/3544548.3581213}
\acmISBN{978-1-4503-9421-5/23/04}

\begin{document}

\title{Understanding How Low Vision People Read Using Eye Tracking}



\author{Ru Wang}
\affiliation{%
   \institution{University of Wisconsin-Madison}
   \city{Madison}
   \state{WI}
   \country{USA}}
\email{ru.wang@wisc.edu}

\author{Linxiu Zeng}
\affiliation{%
   \institution{University of Wisconsin-Madison}
   \city{Madison}
   \state{WI}
   \country{USA}}
\email{lzeng37@wisc.edu}

\author{Xinyong Zhang}
\affiliation{%
   \institution{Renmin University of China}
   \city{Beijing}
   \country{China}}
\email{x.y.zhang@ruc.edu.cn}

\author{Sanbrita Mondal}
\affiliation{%
   \institution{University of Wisconsin-Madison}
   \city{Madison}
   \state{WI}
   \country{USA}}
\email{smondal4@wisc.edu}

\author{Yuhang Zhao}
\affiliation{%
   \institution{University of Wisconsin-Madison}
   \city{Madison}
   \state{WI}
   \country{USA}}
\email{yuhang.zhao@cs.wisc.edu}

\renewcommand{\shortauthors}{}

\begin{abstract}

While being able to read with screen magnifiers, low vision people have slow and unpleasant reading experiences.
Eye tracking has the potential to improve their experience by recognizing fine-grained gaze behaviors and providing more targeted enhancements. To inspire gaze-based low vision technology, we \change{investigate the suitable method to collect low vision users' gaze data via commercial eye trackers and thoroughly explore their challenges in reading based on their gaze behaviors}. With an improved calibration interface, we collected the gaze data of \change{20 low vision participants and 20 sighted controls} who performed reading tasks on a computer screen; low vision participants were also asked to read with different screen magnifiers. We found that, with an accessible calibration interface and data collection method, commercial eye trackers can \change{collect gaze data of comparable quality from low vision and sighted people.} Our study identified low vision people’s unique gaze patterns during reading, building upon which, we propose design implications for gaze-based low vision technology.

\end{abstract}




\maketitle

\section{Introduction}

Reading is a critical daily task for people to access a variety of visual information, from the prescription on the pill bottle to the traffic signs on the side of streets. However, this task could be challenging to low vision people. Low vision is a visual impairment that cannot be corrected by eyeglasses, contact lenses, and other standard treatments~\cite{nei}. There are different low vision conditions, such as central vision loss, peripheral vision loss, night blindness, and blurry vision~\cite{nei}. Different conditions can affect low vision people's reading ability and behaviors in different ways, for example, words may appear distorted to people with central vision loss, and people with severe peripheral vision loss may only see one or two words at a time without being able to scan ahead.

While experiencing visual difficulties, most low vision people prefer using their residual vision~\cite{szpiro2016people,zhao2015foresee}. As a result, they often get close to a reading material to see details, and use low vision aids, such as handheld magnifiers, to magnify the text. On computers and smartphones, they use embedded accessibility features, such as screen magnifiers, enlarged font size, and contrast adjustment, to enhance content visibility. Although these low vision aids can compensate for low visual acuity to some extent, low vision people still face challenges in reading. For example, they experience short visual span and have difficulty in switching lines due to magnification and increased font size~\cite{verghese2021eye, cheong2007relationship, cheong2008relationship}, leading to slow reading. Prior research shows that low vision people read text blocks about 3 times slower than sighted people~\cite{bruggeman2002psychophysics}. More research is needed to understand low vision people's reading behaviors and design technologies to mitigate their reading barriers. 


The advance of eye tracking techniques presents an opportunity to discover low vision people's visual challenges by recognizing their gaze use and enable gaze-based vision enhancement technology. Eye gaze can convey a myriad of information in reading. For example, it indicates readers' visual attention change over time~\cite{vickers2009advances} and elucidates the cognitive processing of words, sentences, and text~\cite{jarodzka2017tracking}. Many works utilize eye tracking to characterize sighted people's reading behaviors~\cite{rayner2010eye, zambarbieri2012eye, rayner1998eye, reichle2003ez} and design gaze-based technology to improve their reading experience~\cite{cheng2015gaze, gowases2011text, maus2020gaze}. 
However, the gaze research for low vision remains nascent. With \change{commercial eye trackers} increasingly integrated into everyday devices (e.g., laptop, AR glasses), there is great potential in leveraging this technology to understand low vision people's detailed gaze behaviors (e.g., when they deviate from a line of text, lose track of the next line, or struggle with recognizing a word), thus designing assistive technologies that provide more targeted support in reading tasks. 


Despite the potential, eye tracking technology could be challenging for low vision people to use. The state-of-the-art commercial eye trackers (e.g., the Tobii eye tracker) recognize the features of users' eyes (e.g., pupil, iris, corneal reflection) and predict their gaze point via a gaze estimation algorithm~\cite{tobiicali}. To account for the individual differences in the eye shape and geometry, a calibration interface is provided that renders visual targets on a display and collects users' eye data when focusing on the targets. Although these calibration and modeling methods can achieve high accuracy for sighted users~\cite{tobiiproacc}, 
they overlook the different visual abilities, eye characteristics, and eye usage strategies that low vision people may have \cite{maus2020gaze}. 
For example, the targets on the calibration interface are small and in a fixed size, which can be invisible to users with low visual acuity, causing low calibration accuracy. \change{Some low vision people had extremely unbalanced visual abilities and thus inconsistent gaze behaviors across the two eyes. Moreover, the appearance of eyes could be changed by particular eye diseases (e.g., cataract), causing recognition failure and high data loss.} 
With these issues, it is unclear whether and how low vision people can use and benefit from the emerging eye tracking technology. 

To close the gap, we seek to explore the \change{potential of collecting high-quality gaze data from low vision people using off-the-shelf eye trackers}, and further leverage a commercial eye tracker to investigate low vision people's detailed gaze behaviors in reading tasks to inspire gaze-based low vision technology.  
To make the eye tracker more usable to low vision people, \change{an adjustable calibration interface and a dominant-eye-based data collection method} were designed and developed. Based on the improved calibration process, 
we conducted a study with \change{20 low vision participants and 20 sighted controls}, who performed reading tasks on a computer screen with an eye tracker collecting their gaze data. 
Since the use of screen magnifiers may affect low vision people's gaze behaviors, we also asked low vision participants to read with different types of screen magnifiers (i.e., lens magnifier, full-screen magnifier) to investigate their gaze patterns and challenges in different magnification modes. 
\change{With this study, we seek to answer four research questions: (RQ1) Can we collect reliable eye gaze data from low vision people using a commercial eye tracker? (RQ2) How do low vision people's gaze behaviors differ from sighted people during reading? (RQ3) How do different visual conditions affect low vision people's gaze behaviors? (RQ4) How do different screen magnification modes affect low vision people's gaze patterns?}

Our research demonstrates that \change{the relatively low-cost, high-availability commercial eye tracker} can be a promising tool for low vision research \change{and collect reliable data from low vision people} if the calibration process is accessibly designed. We also discover unique gaze patterns and visual challenges faced by people with different low vision conditions. For example, low vision participants fixate their gaze more than sighted people at the beginning of each line during line switching. Participants with low visual acuity \change{and limited visual field} show more but shorter fixations than those with relatively high visual acuity \change{and intact visual field. Moreover, when using the lens magnifier,
increasing the height of the magnification window can make line switching easier.} 
Based on our findings, we identify potential improvements for eye tracking technology, and derive design implications for gaze-based assistive technology to support low vision people during reading. 

\section{Related Work}
%
%
%

\subsection{Experience of Low Vision People during Reading}
Reading is a common challenge for low vision people. The reading performance of low vision people has wide variation due to different visual conditions~\cite{legge1985psychophysics}. For example, people with cloudy or blurry vision demonstrate a strong dependence of reading time on word length due to reduced visual span~\cite{legge1997psychophysics}; people with Macular Degeneration which causes central vision loss read more slowly than people with equivalently reduced visual acuity but intact central vision~\cite{krischer1985visual, legge1985psychophysics}.

To leverage their residual vision, low vision people use different magnification methods for reading, such as increasing font size or screen magnifiers.
Today most computers and smartphones support system embedded screen magnifiers~\cite{maczoom, winmag}, which enable users to magnify a particular screen area based on the mouse position on the screen. 
Although screen magnifiers can address the low acuity difficulty to some extent, some usability issues have been reported repeatedly by prior research. For example, Moreno et al.~\cite{moreno2021exploratory} has found that users might lose context when using screen magnifiers because \change{they can only view a partial area of the screen at a time}. 
\change{The reduced field of view caused by magnification also affects low vision users' reading speed, for instance, they recognize fewer letters with one fixation than sighted people} \cite{verghese2021eye, cheong2007relationship, cheong2008relationship}.
Moreover, moving the mouse around to magnify different areas on the screen demands high spatial visualization skills, \change{increasing users' cognitive load during reading}~\cite{hallett2015reading, szpiro2016people}. \change{This issue makes it particularly difficult to reposition the magnifier to locate the next line, especially for users with visual field loss}~\cite{ahn1995psychophysics, verghese2021eye}. 

With the limitations of screen magnifiers, low vision people's reading experience is largely diminished. Prior research shows that they read text blocks 3.2 times slower than sighted people \cite{hallett2015reading, bruggeman2002psychophysics}. As such, more technologies (e.g., gaze-based technology) are needed to mitigate the reading barriers for low vision people. 




\subsection{Understanding Human Gaze via Eye Tracking}
Eye gaze conveys a variety of information. Understanding human gaze can help elucidate how people complete different visual tasks and \change{provide insights to augment low vision people's reading experience}. 

\subsubsection{How does eye tracker work?}
\begin{figure}
  \includegraphics[width=0.4\textwidth]{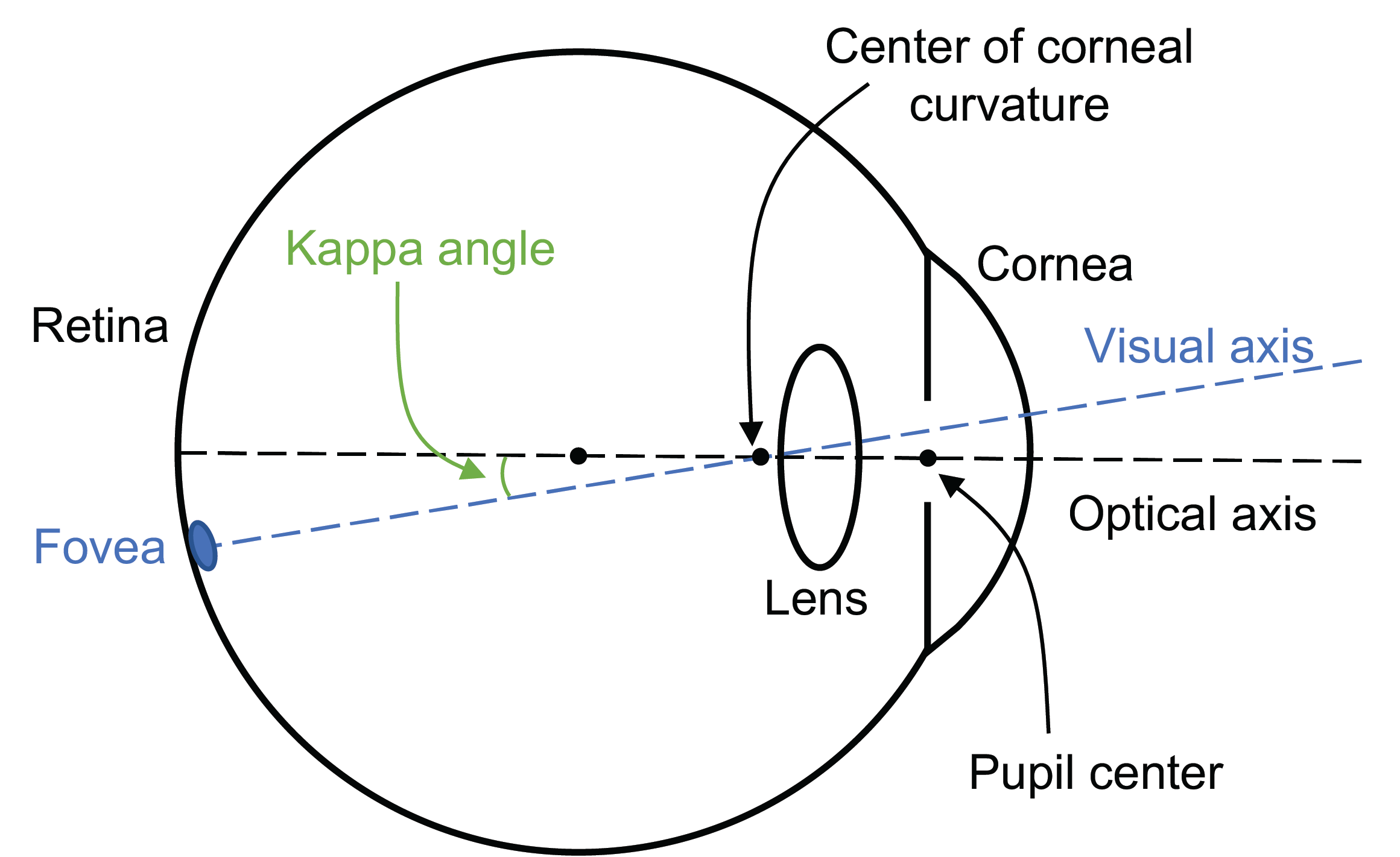}
  \caption{Model of human eye ball. The optical axis connects the center of corneal curvature and the pupil center; the visual axis connects the fovea and the center of corneal curvature. Kappa angle is the angular deviation between the optical and visual axis~\cite{kar2017review}.}
  \label{fig:eye}
  \Description{The figure shows the structure of a human eye ball. There are four important components of a human eye in the figure: cornea, pupil, lens, retina and the fovea on retina. The optical axis connects the center of corneal curvature and the pupil center; the visual axis connects the fovea and the center of corneal curvature. Kappa angle is the angular deviation between the optical and visual axis.}
    \vspace{-2ex}
\end{figure}

\change{Commercial eye trackers are commonly used in most eye-tracking research due to its high accuracy and ease-of-use~\cite{zhang2021eye}.} 
Screen-based eye tracking involves capturing the user's face and eye region, detecting pupils, and mapping the user's gaze to the computer screen coordinate~\cite{kar2017review}. However, the gaze direction cannot be directly determined based on the pupil position: the area of retina with highest acuity is called fovea~\cite{kar2017review, guestrin2006general}, and the visual axis connecting fovea and the center of corneal curvature determines the true gaze direction; the angular offset between the optical axis (which can be determined by the user's pupil and head position) and visual axis---which is called kappa angle---is user dependent~\cite{kar2017review, zhu2007novel} (Fig.\ref{fig:eye}). Therefore, a calibration needs to be completed to solve the kappa angle, so that an accurate mapping between eye images to gaze directions can be obtained. 
The calibration for eye tracker is usually performed by asking the user to gaze at multiple small targets distributed over the screen for a certain amount of time~\cite{kar2017review}. The eye tracker can then collect the user's eye images corresponding to these target points and estimate the kappa angle. 

However, the same procedure might not work for low vision users~\cite{shanidze2020eye}. For example, some low vision people's visual acuity might be too low to see the calibration targets clearly \cite{murai2010eye}, leading to inaccurate calibration results; 
\change{some may have irregular eye appearance that prevents pupil recognition (e.g., Coloboma~\cite{pagon1981ocular})}.
\change{It is thus important to explore how to refine the calibration and data collection process to collect high-quality gaze data from low vision users.} 

\subsubsection{Understanding the gaze behaviors of sighted people.} A myriad of prior research has leveraged eye tracking technologies to understand sighted people's gaze behaviors \cite{rayner1998eye, rayner2010eye, zambarbieri2012eye}. Two eye movement events are typically used to measure and interpret gaze behaviors, including \textit{fixation} (i.e., gaze pauses over informative regions of interest) and \textit{saccade} (i.e., rapid, ballistic movements between fixations) \cite{salvucci2000identifying}. In reading scenarios, fixations and saccades encode abundant information regarding the process of perception and comprehension~\cite{hallett2015reading, rayner1998eye, rayner2010eye, jarodzka2017tracking}.
 For example, the duration of a fixation indicates the difficulty of perceptual or cognitive processing during reading since visual information can only be perceived during fixations~\cite{zambarbieri2012eye}. The main function of saccades is to move a new block of text into foveal vision where visual acuity is the highest, because reading with parafoveal or peripheral vision is difficult~\cite{rayner1998eye, verghese2021eye}. Saccades can be further categorized according to their roles~\cite{bax2013cognitive, rayner2012psychology, rayner1998eye}. Forward saccades happen when reading onwards in left-to-right languages. Regressive saccades are conducted to revisit words covered by previous forward saccades in order to enhance comprehension~\cite{hallett2015reading}. Return sweeps are to switch focus from the end of one line to the beginning of the next. Interestingly, during a return sweep, readers often undershoot and make small corrective movements to reach the accurate location, which also leads to regressive saccades~\cite{rayner1998eye}. Besides, prior work \cite{bowers2017microsaccades, engbert2003microsaccades} has also studied the function of microsaccades (i.e., very small saccades) and suggests that microsaccades are used to correct previously made long saccades and gain additional information about adjacent words~\cite{bowers2017microsaccades}.



\subsubsection{Eye tracking research for low vision.} Although previous research on eye tracking technology and understanding human gaze is fruitful, very few efforts have been made to investigate the eye movement of low vision people while performing daily tasks. 

Some researchers in optometry and vision science have started investigating low vision people's gaze patterns using eye trackers. Most works focus on people with central vision loss, who need to adopt an eccentric retinal location as a substitute for the fovea, also known as the preferred retinal locus (PRL). Such research explores how the use of PRL affects low vision people's visual tasks by tracking their eye movement \cite{prahalad2020asymmetries, verghese2021eye, renninger2011recalibration, kwon2013rapid}. For example, Rubin and Feely \cite{rubin2009role} have used an SMI high-speed video tracker to evaluate the fixations and saccades of 40 people with age-related macular degeneration, showing that participants' reading performance were significantly associated with the fixation stability, proportion of regressive saccades, and length of forward saccades. Prahalad and Coates \cite{prahalad2020asymmetries} use an EyeLink 1000 Plus eye tracker (the approximate price is around \$30,000 \cite{eyelink}) to explore how people with central vision loss choose their PRL. They have recruited six sighted people who experience simulated scotoma. Participants are trained to use different artificial PRL (left, right, inferior) to read. The study shows that a right PRL causes \change{longer saccades} and more directional switches, indicating that people with central vision loss may avoid right PRL due to an oculomotor reason. Moreover, Bullimore and Bailey \cite{bullimore1995reading} use a specialized infrared scleral reflection device~\cite{brown1977clinically} to track the eye movement of both sighted participants and participants with age-related maculopathy. They find that participants with central vision loss demonstrate a large number of regressive saccades because the scotoma distorts or occludes part of the words. 

However, research in vision science usually leverages clinical or research eye trackers, which are highly expensive and not commonly available for everyday use. \change{Some researchers even created specialized eye trackers for their research} \cite{brown1977clinically, robinson1963method}. Moreover, none of the above research detailed the specific calibration and data collection process for low vision participants and the potential challenges they may encounter. 

There has been very limited research in HCI that investigates gaze-based technology for low vision people using eye trackers. Murai et al. \cite{murai2010eye} have used Mobile Eye, a device unit with an eye-camera and a scene-camera to recognize people's gaze point in the environment. They design a calibration interface with large and high contrast letters for low vision people and estimate their gaze circles (i.e., an area the low vision user focuses on) instead of gaze points. They find that while it is possible to calibrate for people with severe low vision, the eye tracker could not be used by people with nystagmus (i.e., involuntary eye movement). However, this research has been conducted a decade ago. With more advanced commercial eye trackers deployed to the market, it is necessary to revisit this problem and investigate how to better leverage the state-of-the-art commercial eye trackers for low vision research. More recently, Masnadi et al. \cite{masnadi2020vriassist} have designed VRiAssist, a gaze-based vision enhancement tool in virtual reality that dynamically follows the user's gaze and generates visual corrections, thus helping low vision people see better in a virtual space. However, no information is provided regarding what eye tracking technology is used and how they calibrate and collect gaze data from low vision users. Moreover, Maus et al. \cite{maus2020gaze} have designed a gaze-guided magnification with a commercial GazePoint GP3 eye tracker to enable low vision people to control the on-screen magnifier with their gaze. They demonstrate that the current video-based eye tracking technique does not consider low vision people and result in a large amount of data loss. 

While researchers have recognized the potential of eye tracking and gaze-based vision enhancements for low vision people \cite{masnadi2020vriassist, maus2020gaze, murai2010eye}, it is still unclear whether and how low vision people can leverage such technology in their daily life. Our research will fill this gap and explore the feasibility of off-the-shelf eye trackers for low vision people by refining the calibration and data collection process, as well as investigating their gaze behaviors during reading tasks to inspire gaze-based technology.




\section{Method}
\label{methods}
Our goal is to \change{explore the potential of using a commercial eye tracker to collect high-quality gaze data from low vision people. With suitable eye tracking process, we further investigate low vision people's unique gaze patterns compared to sighted people as well as their different gaze behaviors due to different visual conditions and magnification modes. We thus design a study to assess four groups of hypotheses based on the four research questions in Introduction:}
\begin{itemize}[leftmargin=2em]
    \item[\textbf{H1}] \textbf{With improved calibration and data collection process, the commercial eye tracker can achieve comparable data quality from low vision people and sighted people.}

    \begin{itemize}[align=left, leftmargin=0.25cm]
        \item[\change{H1.1}] \change{There is no difference in the gaze recognition accuracy between low vision and sighted people.}
        \item[\change{H1.2}] \change{There is no difference in the percentage of data loss between low vision and sighted people.}
    \end{itemize}

    \item[\textbf{H2}] \textbf{Low vision people have different gaze behaviors from sighted people when reading.}

    \begin{itemize}[align=left, leftmargin=0.25cm]
        \item[\change{H2.1}] \change{Low vision people show different fixation patterns from sighted people.}
        \item[\change{H2.2}] \change{Low vision people show different saccade patterns from sighted people.}
        \item[\change{H2.3}] \change{Low vision group shows different line searching behaviors from sighted group.}
    \end{itemize}

    \item[\textbf{H3}] \textbf{Different visual abilities (i.e., visual acuity, visual field) affect low vision people's gaze behaviors differently.}

    \begin{itemize}[align=left, leftmargin=0.25cm]
        \item[\change{H3.1}] \change{Different visual abilities affect low vision people's fixation patterns differently.}
        \item[\change{H3.2}] \change{Different visual abilities affect low vision people's saccade patterns differently.}
        \item[\change{H3.3}] \change{Different visual abilities affect low vision people's line searching behaviors differently.}
    \end{itemize}

    \item[\textbf{H4}] \textbf{Different screen magnification modes (i.e., enlarged font, lens magnifier, full-screen magnifier) affect low vision people's gaze behaviors in reading differently.}

    \begin{itemize}[align=left, leftmargin=0.25cm]
        \item[\change{H4.1}] \change{Different magnification modes affect low vision people's fixation patterns differently.}
        \item[\change{H4.2}] \change{Different magnification modes affect low vision people's saccade patterns differently.}
        \item[\change{H4.3}] \change{Different magnification modes affect low vision people's line searching behaviors differently.}
    \end{itemize}
    
\end{itemize}

\subsection{Participants}
We recruited \change{20 low vision and 20 sighted participants to compare their gaze patterns during reading tasks.} Our low vision participants included \change{14 female and 6 male, whose ages ranged from 19 to 86 ($Mean=58.3, SD=22.1$)}. \change{Seven} participants were legally blind. Their visual conditions are detailed in Table \ref{tab:lv_dem}. \change{ Besides low vision, participants did not have any other health conditions that may cause reading difficulties.} \change{We recruited low vision participants from a local low vision clinic. We also posted recruitment messages on the student jobs website of our university.} 
When a potential low vision participant contacted the research team, we conducted a brief interview via phone, email, or text to ensure they were eligible to our study. A participant was eligible if they were at least 18 years old and had low vision. \change{Only one participant (Hannah) wore glasses during the study.} Participants were compensated \$20 per hour and were reimbursed for travel expenses. 

Our sighted participants included \change{7 female and 13 male, with ages ranging from 21 to 51 ($Mean=31.1, SD=9.5$)}. All participants' visual acuity in the better eye was no worse than 20/40. \change{Three of them wore eye glasses, and two wore contact lenses.} 
They were compensated \$10 per hour for participating in the study.

\begin{table*}[t]
\footnotesize
\centering
\begin{tabular}{C{1cm}C{0.8cm}C{2.9cm}C{1cm}C{1.5cm}C{2.7cm}C{2.9cm}}
\toprule
\textbf{Pseudonym}&\textbf{Age/{\newline}Gender}&\textbf{Diagnosed{\newline}Condition}&\textbf{Legally {\newline} Blind?} &\textbf{Visual {\newline} Acuity}&\textbf{Visual Field}&\textbf{Accessibility {\newline} Features Used}\\
\midrule
Tim & 70/M & Macular Degeneration & N & R: 20/160 {\newline} L: 20/128 & Distorted central vision & Large font\\ 
\hline
Judy & 60/F & Spinal Meningitis & Y & R: 20/400 {\newline} L: 20/2200 & Bottom peripheral vision loss & Lens magnifier\\ 
\hline
Bella & 68/F & Dry Macular Degeneration & N & R: 20/100 {\newline} L: 20/128 & Central vision loss & Bold and large font\\ 
\hline
Mark & 78/M & Cone Dystrophy & N & R: 20/100 {\newline} L: 20/48 & Central vision loss & Large screen, full-screen magnifier {\newline} Bold and large font \\ 
\hline
Kate & 60/F & Ocular Melanoma (Radiation Retinopathy) & N & R: 20/64 {\newline} L: 20/48 & Left peripheral vision loss on left eye & High contrast, bold and large font, low brightness \\ 
\hline
Robin & 33/M & Retinitis Pigmentosa & Y & R: 20/40 {\newline} L: 20/40 & Peripheral and part of central vision loss & N/A \\ 
\hline
Hailey & 19/F & Retinal Scarring (left eye) & N & R: 20/12 {\newline} L: 20/100 & Central vision loss on left eye & Low brightness \\ 
\hline
Lucy & 61/F & Non-arterial Ischemic Optical Neuropathy, Diabetic Retinopathy & Y & R: <20/160 {\newline} L: 20/48 &  Peripheral and part of central vision loss & Screen reader \\ 
\hline
Mary & 83/F & Macular Degeneration, Charles Bonnet Syndrome & N & R: 20/128 {\newline} L: 20/100 &  Distorted central vision & Full-screen magnifier \\ 
\hline
Caroline & 19/F & Nystagmus (right eye) & N & R: <20/160 {\newline} L: 20/40 & Right eye lagging behind & N/A \\ 
\hline
Hannah & 65/F & Glaucoma & N & R: 20/100 {\newline} L: 20/40 & Peripheral vision loss & Large font \\ 
\hline
Maeve & 20/F & Cataract (right eye) & N & R: <20/160 {\newline} L: 20/24 & Blurry and darker on right eye  & N/A \\ 
\hline
\change{May} & \change{77/F} & \change{Retinitis Pigmentosa} & \change{Y} & \change{R: 20/64 {\newline} L: 20/40} & \change{Peripheral vision loss} & \change{Bigger and Brighter~\cite{biggerbrighter}} \\ 
\hline
\change{Piper} & \change{56/F} & \change{Retinitis Pigmentosa} & \change{Y} & \change{R: 20/128 {\newline} L: 20/160} & \change{Peripheral vision loss} & \change{ Large font, full-screen magnifier, Bigger and Brighter~\cite{biggerbrighter}, Envision App~\cite{envision}} \\ 
\hline
\change{Diego} & \change{74/M} & \change{Wet Macular Degeneration} & \change{N} & \change{R: 20/100 {\newline} L: <20/160} & \change{Central vision loss on left eye} & \change{ Large font, invert color } \\ 
\hline
\change{Danilo} & \change{58/M} & \change{Diabetic Retinopathy} & \change{Y} & \change{R: <20/160 {\newline} L: <20/160} & \change{Blurry central vision on left eye, only light and movement perception on right eye} & \change{Large font, invert color, full-screen magnifier} \\ 
\hline
\change{Ryan} & \change{86/M} & \change{Macular Degeneration} & \change{N} & \change{R: 20/64 {\newline} L: 20/80} & \change{Blurry central vision} & \change{Large font, full-screen magnifier} \\ 

\hline
\change{Marilyn} & \change{70/F} & \change{Glaucoma, Fuchs Heterochromic Uveitis, Coloboma} & \change{Y} & \change{R: <20/160 {\newline} L: <20/160} & \change{Blurry central vision} & \change{Large font, full-screen magnifier, invert color, touch screen for PC} \\ 
\hline
\change{Julia} & \change{80/F} & \change{Macular Degeneration} & \change{N} & \change{R: 20/80 {\newline} L: <20/160} & \change{Central vision loss} & \change{Large font} \\ 

\hline
\change{Fiona} & \change{29/F} & \change{Congenital Glaucoma, Cateract} & \change{N} & \change{R: 20/160 {\newline} L: <20/160} & \change{Peripheral vision loss} & \change{Large and bold font, invert color } \\

\bottomrule
\end{tabular}
\caption{\change{Demographic information of the 20 low vision participants. The visual acuity was measured in our lab except Judy.}}
\label{tab:lv_dem}
  \vspace{-5ex}

\end{table*}

\subsection{\change{Improved Gaze Calibration and Collection Process}} 
\label{sec: calibration}

We used a Tobii Pro Fusion (120Hz) eye tracker~\cite{tobiiprofusion} in the study. To better collect low vision people's data, we refined the gaze calibration and collection process as below.

\subsubsection{\change{Adjustable} Calibration Interface \change{for Low Vision}.} 
\begin{figure*}[h]
  \includegraphics[width=0.7\textwidth]{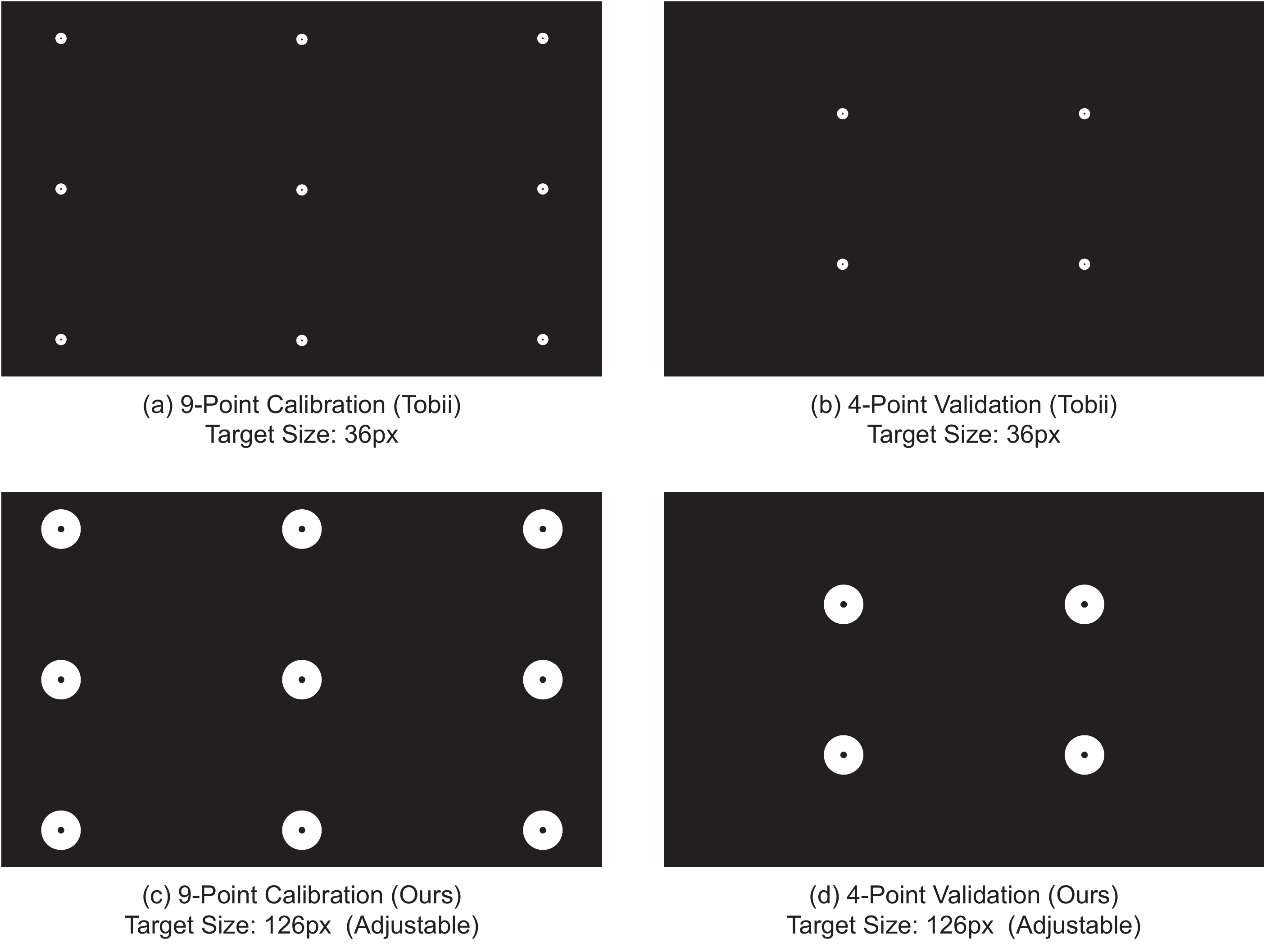}
  \caption{The gaze calibration and validation interfaces. (a) The calibration interface of Tobii Pro Lab (TPL); (b) The validation interface of TPL; (c) Our adjustable calibration and (d) validation interface.}
  \label{fig:cali_ui}
  \Description{Four screenshots labeled (a), (b), (c) and (d) show the gaze calibration and validation interfaces. Image (a) shows the 9-point calibration interface of TPL with 9 calibration targets (target size: 36px) distributed evenly across the screen. (b) shows the 4-point validation interface of TPL with 4 validation targets (target size: 36px) distributed evenly across the middle 40\% area of the screen. (c) shows our adjustable 9-point calibration interface with 9 calibration targets (target size: 126px in the image, adjustable) distributed evenly across the screen. (d) shows our adjustable 4-point validation interface with 4 validation targets (target size: 126px in the image, adjustable) distributed evenly across the middle 40\% area of the screen.}
    \vspace{-2ex}
\end{figure*}

To collect gaze data more accurately, a calibration is needed to mitigate individual differences. 
Tobii Pro Lab (TPL, Tobii's default calibration interface)~\cite{tobiiprolab} adopts a 9-point calibration process: nine targets (white solid circle with a black dot in the center) are shown one by one on the screen and users are asked to fixate on each target until it disappears (Fig. \ref{fig:cali_ui}a). A 4-point validation process is then conducted by asking the user to fixate at another four targets to evaluate the calibration (Fig. \ref{fig:cali_ui}b). 
Although TPL provides some customization to the target color and background color, the target size is always fixed at 36px. 

We refined TPL's calibration interface to make it more usable to low vision people. We kept the high contrast setting (white target on black screen) and further made the target size adjustable (Fig. \ref{fig:cali_ui}c-d), because a small target may not be visible to participants with low visual acuity. \change{The target size ranges from 36px to 256px. With this interface, we can enlarge the targets to a visible size for low vision participants to facilitate a more accurate calibration.} 


\subsubsection{\change{Dominant-eye-based} Data Collection.} 
\change{Conventional gaze collection focused on binocular data collection (i.e., mean gaze position of the two eyes) since sighted people usually show same gaze behaviors in the two eyes. However, some low vision people have distinct visual abilities and thus inconsistent gaze behaviors across the two eyes. We thus collected low vision people's gaze data by considering their dominant eye. This was inspired by Maus et al.~\cite{maus2020gaze}, where they collected low vision participants' gaze data based on two eyes, resulting in high data loss. They thus recommended using monocular data collection to mitigate this issue. 
We collected a participant's gaze from the dominant eye if (1) the participant indicated that they had a dominant eye, and (2) they demonstrated inconsistent gaze behaviors in two eyes (e.g., not looking at the same direction) or one of their eyes was not trackable (e.g., unrecognizable pupil).}
Otherwise, we used binocular data collection, which is the same as the method for sighted participants. 

\subsection{Apparatus: \change{Study Interfaces and Environment}}
Our study was conducted in a well-lit lab. Participants were seated in front of a computer display (24-inch, 1920x1200 resolution) with a Tobii eye tracker attached at the bottom of the screen (Figure \ref{fig:setting}). 
We built a web-based interface using React~\cite{react}, which included the improved gaze calibration interface for low vision participants (section \ref{sec: calibration}) and a reading task interface with different magnification modes to collect gaze data. We used the Tobii Pro SDK~\cite{tobiiprosdk} in Python to retrieve gaze data from the eye tracker. To enable communication between the eye tracker and the reading interface, we built a Flask~\cite{flask} server in Python.

\subsubsection{Reading Task Interface} 
We implemented a reading task interface to collect participants' gaze data and audio data when they were reading aloud. We presented the reading materials in San Francisco font (Sans-serif). By default, the font size was 12pt (16px). The text spacing was set based on the Web Content Accessibility Guidelines (WCAG) 2.1 ~\cite{wcag}. 

For low vision participants, the interface provided three magnification modes, including the regular mode with increased font size (Fig. \ref{fig:reading_ui}a), the lens magnifier (Fig. \ref{fig:reading_ui}b), and the full-screen magnifier (Fig. \ref{fig:reading_ui}c). 


The regular mode was the same as the reading interface presented to sighted participants, but low vision participants were allowed to customize the font size (up to 16 times of default font size) \change{to ensure the readability of the reading content.}
The lens magnifier (Fig. \ref{fig:reading_ui}b) and full-screen magnifier (Fig. \ref{fig:reading_ui}c) \change{simulated commonly-used screen magnifiers embeded in} Windows 10~\cite{winmag}. The lens magnifier allowed the user to magnify \change{a rectangular area of the screen around the mouse cursor; the user could adjust the width and height of the lens magnifier and move the mouse to magnify different areas of the screen}. The full-screen magnifier magnified the whole screen; the user was allowed to pan around with their mouse to reveal different areas of the screen. The user could adjust the zoom level for both magnifiers. For all three magnification modes, low vision participants could also adjust the font weight (bold vs. regular) and text color (white-on-black vs. black-on-white) to \change{simulate the reading setup they used in daily life} (Fig. \ref{fig:reading_ui}d). 

\subsubsection{Visual Function Tests}
\label{sec:vision test}
To identify participants' visual ability, we tested their visual acuity and field of view.

\textbf{Visual acuity test.} We printed a letter-size ETDRS 1 and ETDRS 2 logMAR chart~\cite{ferris1982new} to test the visual acuity of participants' left eye and right eye, respectively. The logMAR chart measures visual acuity (ranges from 20/8 to 20/160 at 5-feet distance) by showing rows of letters in decreasing size. 

\begin{figure*}[h]
  \includegraphics[width=\textwidth]{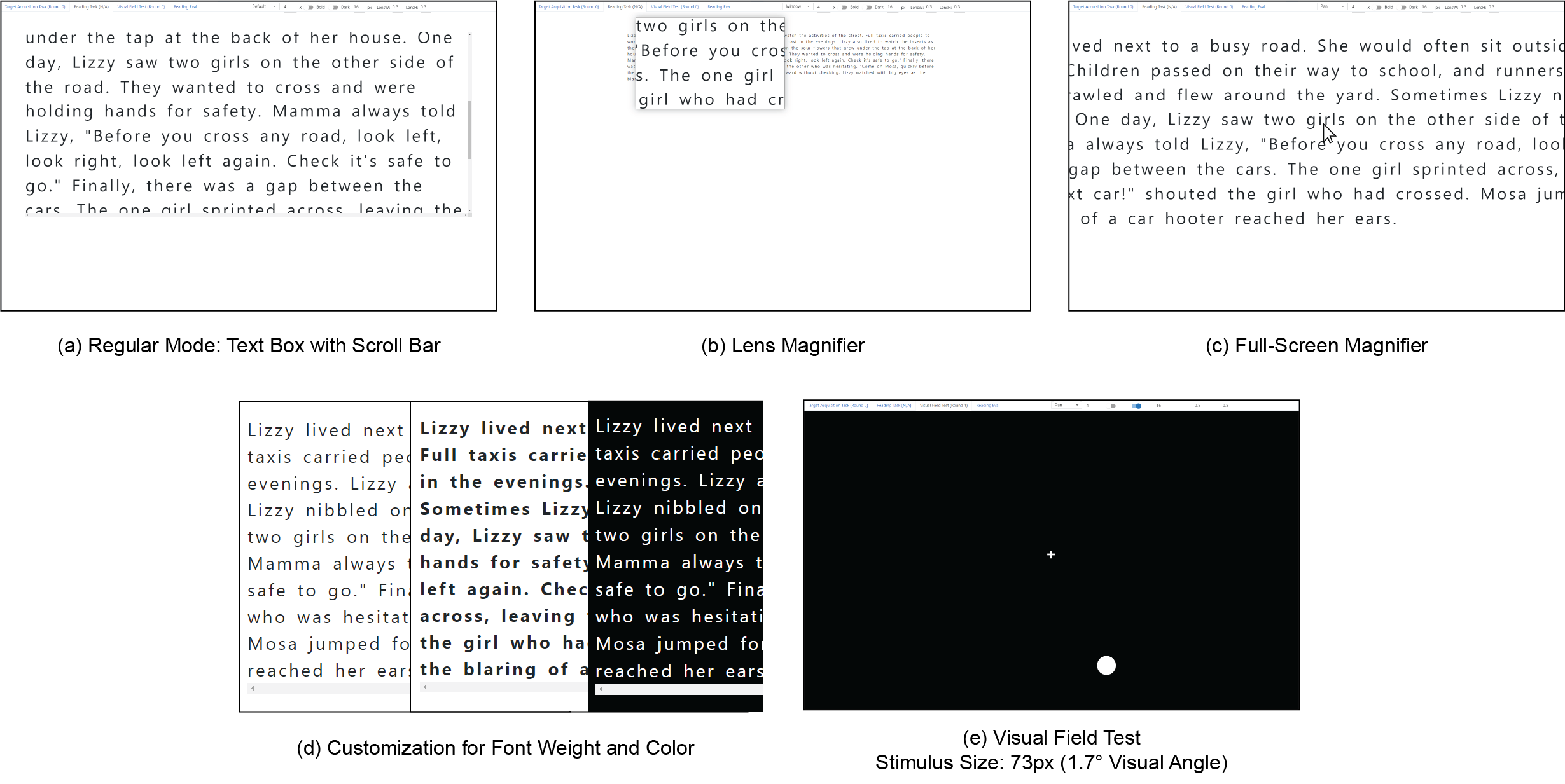}
  \caption{The reading interface and visual field test: (a) Regular mode with increased font size; (b) Lens magnifier; (c) Full-screen magnifier; (d) Adjustable font weight and color for low vision participants in reading tasks; (e) The visual field test interface. }
  \label{fig:reading_ui}
  \Description{Five screenshots labeled (a), (b), (c), (d) and (e) show the reading task and visual field test interfaces. Image (a) shows an example of the regular mode (a text window with large font in the middle of the screen). (b) shows an example of the lens magnifier (the rectangular window magnifies the text underneath it). (c) shows an example of the full-screen magnifier (the whole text is magnified and exceeds the screen). (d) shows examples of the adjustable font weight and color for low vision participants in reading tasks. From left to right, it shows the regular font style, bold font, and inverted color (white-on-black). (e) shows the visual field test interface with a fixation target in the center of the screen and a stimulus of size 73px near the bottom of the screen.}
    \vspace{-2ex}
\end{figure*}

\textbf{Visual field test.} We also built a visual field test interface to roughly identify the areas of vision loss for low vision participants. \change{The test area was limited to the screen size. Although this was not a standard visual field test that covered participants' full field of view, it indicated the influence of low vision participants' visual field loss on their ability to perform visual tasks on the screen.}
The design was based on the Octopus perimeter~\cite{racette2016visual} but simplified. Participants were instructed to look straight ahead on a fixation target in the center of the screen all the time with both eyes open (Fig. \ref{fig:reading_ui}e). 
During the test, visual stimuli appeared randomly anywhere on the screen; one stimulus appeared at a time. Participants were asked to press the SPACE key as soon as they noticed a stimulus in their peripheral visual field. 

The stimulus was a flash of white solid circle on the screen that lasted 0.1s.
\change{The diameter of the stimulus was 1.7\textdegree~ viewed at 65cm from the screen, which was the standard stimulus size for low vision people in Octopus perimeter. 
The stimuli spanned a 40\textdegree~ $\times$ 20\textdegree~ area viewed at 65cm from the screen, thus including central vision and part of near peripheral vision. 
The stimuli were evenly distributed on a 5\textdegree~ spacing grid, resulting in 45 stimuli (9 $\times$ 5) in total.}

\subsection{Procedure}
We conducted a single-session study that lasted 1 to 2 hours. Participants were invited to our lab for the study. \change{The study included four phases below:}

\change{\textbf{\textit{Initial Interview \& Visual Acuity Test.}}} After obtaining the participants' consent, we conducted a brief semi-structured interview. For all participants, we asked about their demographics. For low vision participants specifically, we asked about their visual condition, dominant eye, experience with assistive technology, and any adaptations to their vision on their daily devices. 
We then measured their visual acuity (both with and without correction if they wore eyeglasses) using the ETDRS eye charts (Section \ref{sec:vision test}). Participants sat at 5 feet in front of the chart and read from the top line on the chart. They were asked to read the letters on chart 1 with left eye covered then read chart 2 with right eye covered. \change{We regarded a line as visible to a participant if they correctly identified more than three out of five letters on that line. We recorded the lowest visible line for each participant.} 

\begin{figure}[h]
  \includegraphics[width=0.3\textwidth]{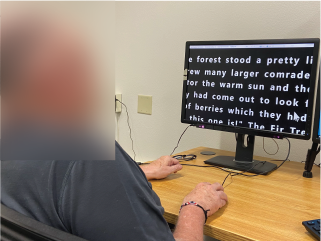}
  \caption{A low vision participant performing reading task with full-screen magnifier in front of a screen}
  \label{fig:setting}
  \Description{The figure shows a low vision participant performing reading task with large bold font and inverted color using the full-screen magnifier on a screen with eye tracker mounted at the bottom.}
    \vspace{-2ex}
\end{figure}

\textbf{\textit{Gaze Calibration and Validation.}}
\change{We conducted gaze calibration for participants with our improved calibration interface (Section \ref{sec: calibration}).} Participants were instructed to sit in a chair in front of a computer monitor. We asked them to sit straight with their back touching the back of the chair to keep a stable position throughout the study. Participants adjusted their position so that the horizontal distance between their eyes and the monitor was about 65cm---the optimal distance for the eye tracker to work. We adjusted the height of the chair so that participants' eye level was at the center of the screen. After reaching the suitable position, we asked participants to try their best to keep their head and body still across the tasks (Fig \ref{fig:setting}). 
We increased the calibration target size for low vision participants until they were able to see the center of the target without squinting, while sighted participants used the default target size. Participants completed a 9-point calibration and 4-point validation \change{under the instructions from the research team}. We asked participants to repeat the calibration if the validation result was larger than one degree or if the research team noticed them not focusing on the targets \change{or blinking} during data collection. 

\textbf{\textit{Gaze Collection during Reading.}}
After calibration, participants were asked to complete multiple trials of reading tasks in front of the eye tracker. Both sighted and low vision participants read two passages in the \textbf{regular mode} (Fig. \ref{fig:reading_ui}a). 
\change{While all sighted participants could read the default font comfortably, the text was not readable to all low vision participants. As such, to stimulate participants' natural gaze behaviors, we allowed low vision participants to adjust the font size, weight, and color to the settings they used in daily life. To explore the effect of different screen magnifiers on low vision participants' gaze patterns, we also asked them to read in the other two screen magnification modes (two passages per condition): the \textbf{lens magnifier} and \textbf{full-screen magnifier}. Participants were also allowed to adjust the window size in the lens magnifier to their preferred size. 
We counterbalanced the order of the three reading modes across low vision participants using Latin Square~\cite{bradley1958complete}.
Four low vision participants did not need magnification due to relatively good visual acuity, they thus only read in the regular mode with default font size (16px) like sighted participants.}




We collected participants' gaze data in each reading task. Before formal data collection, we also asked participants to practice with each magnification mode until they felt comfortable reading with it.
Since we were interested in participants' gaze behaviors at the word and sentence level (i.e., whether participants can locate and recognize each word successfully ~\cite{jarodzka2017tracking}), we instructed them to read aloud the passages to match with the eye movement data. Participants were asked to read as accurately and quickly as possible.

We selected six passages with the similar difficulty level from CLEAR corpus, a corpus including about 5000 text excerpts for readability assessment~\cite{crossley2022large}. The corpus provided readability score for each text excerpt. We selected passages as follows: we first gathered the excerpts in sixth grade level reading difficulty using Flesch Reading Ease~\cite{flesch1948new}. We then calculated the cosine similarity of each excerpt based on their Flesch-Kincaid Grade Level~\cite{kincaid1975derivation}, Automated Readability Index~\cite{kincaid1975derivation}, SMOG~\cite{mc1969smog}, and Google Word Count, resulting in 26 similar-difficulty excerpts. Among them, we selected 12 most similar passages with neutral content (non-politically sensitive) verified by our research team. The mean word count of all selected passages was 185.1 ($SD=8.1$). We then selected six passages as the default passages in our study and the presentation order was randomized. The rest passages were used as back up passages to handle particular situations, such as data collection failures due to system errors.

\begin{figure*}[h]
  \includegraphics[width=\textwidth]{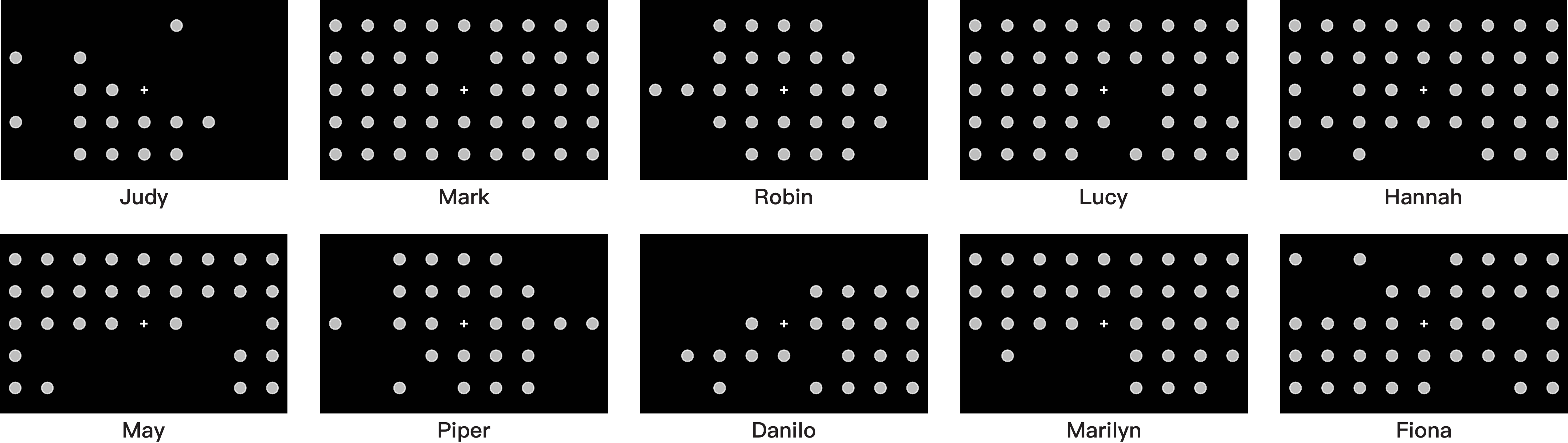}
  \caption{\change{Visual field test result of 10 participants who had limited visual field}}
  \label{fig:field}
  \Description{
This figure shows the visual field test result of participants who had limited visual field. On the first row, from left to right, it shows the visual field test result of Judy, Mark, Robin, Lucy and Hannah. On the second row, from left to right, it shows the visual field test result of May, Piper, Danilo, Marilyn and Fiona. Each result image shows the position of the stimuli that participants were able to see during the test.
}
    \vspace{-2ex}
\end{figure*}

\textbf{\textit{Exit Interview \& Visual Field Test.}}
We ended our study with an interview, asking about participants' reading experiences based on our observation, for example, their difficulty of recognizing specific words, how they tracked the line they were reading, and how they located the next line. We also probed suggestions for improving their reading experiences. Lastly, we tested low vision participants' on-screen field of view with our field of view test interface (Fig. \ref{fig:reading_ui}e). We conducted this test by the end because the visual field test could potentially strain participants' eyes, affecting their reading performance. Fig. \ref{fig:field} shows the test results of the 10 participants with limited visual field. 

\subsection{Analysis}
Our study collected both quantitative and qualitative data. \change{We first describe our quantitative analysis for the four groups of hypotheses and then the qualitative analysis method used for the interview data.}

\subsubsection{\change{Assessing gaze data quality between low vision and sighted people (H1).}} 
\label{h1}
We first \change{validated the calibration and assessed the quality of the data collected from the eye tracker}. We had two measures: (1) \textbf{\textit{gaze recognition accuracy in four-point validation}}, the mean visual angle difference between the target position and the estimated gaze location \change{in the four-point validation (Fig. \ref{fig:cali_ui}d), which was used to validate the calibration~\cite{tobiiproacc}}, and (2) \textbf{\textit{data loss}}, \change{the ratio of the number of invalid gaze points due to eye recognition failure to the total number of gaze points we collected in each recording.}

We \change{used sighted people's data as the standard} and compared the two measures between low vision and sighted people. \change{We thus had one between subject factor \textit{\textbf{Vision}} with two levels---\textit{Sighted} and \textit{LowVision}. Since neither of the measures was normally distributed based on the Shapiro-Wilk test, we used a one-way Aligned Rank Transform for Nonparametric Factorial ANOVA (ART)~\cite{wobbrock2011aligned} to model the impact of \textit{Vision} on both gaze recognition accuracy and data loss.}
Moreover, we aligned participant's audio data and gaze data by matching the data timestamp, in order to confirm whether the collected gaze behaviors matched participants' reading behaviors. 

\subsubsection{\change{Comparing gaze behaviors between low vision and sighted people (H2).}} 
\label{h2}
We then analyzed low vision and sighted participants' gaze data during the reading tasks. We used REMoDNaV~\cite{dar2021remodnav}, an eye-movement \change{event} classification library to recognize eye-movement events. We specify the measures used in our analysis:

(1) \textbf{Fixations} are the pauses over informative regions of interest~\cite{salvucci2000identifying} where eyes are relatively stationary. We measure participants' \textbf{fixation number} and \textbf{mean fixation duration} during reading. 

(2) \textbf{Saccades} are the rapid, ballistic movements between fixations \cite{salvucci2000identifying, verghese2021eye}. We investigate different types of saccades, including \textbf{forward saccades} (i.e., left-to-right saccades when reading onwards) and \textbf{regressive saccades} (i.e., right-to-left saccades that bring eyes back to previously read content). We measure both \textbf{saccade number} and \textbf{saccade length} to understand participants' scanning and revisiting behaviors. Since low vision participants read in different font sizes, we normalize the saccade length by dividing \change{the angular length of a saccade with the angular width of the font based on} the font size selected by each participant. 

(3) \textbf{Revisitation} describes the behavior that a participant revisits prior content during reading. Since it is a corrective movement for a long forward saccade to send the eyes to the accurate position of a piece of text, we define a revisitation pattern as a regressive saccade following a forward saccade and a fixation. We then define \textbf{revisitation rate} as the number of revisitations divided by the total number of forward saccades, and use it to evaluate the frequency that participants corrected their visual scanning.

(4) We measure a participant's line switching behaviors via \textbf{the mean number of searched lines}, meaning the average number of lines a participant searches before they successfully locate the correct next line in a passage. Suppose $N$ denotes the total number of lines in a passage and $nLS_{i}$ represents the number of searched lines when switching to the $i^{th}$ line. We define the mean number of searched lines as $\frac{1}{N-1}\sum_{i=2}^N nLS_i $.



\change{We compared low vision and sighted participants' gaze behaviors. Given that screen magnifiers can largely change low vision participants' reading behaviors, our comparison only focused on the regular mode since the interaction involved in this mode is similar to sighted people's reading experience. For line switching analysis, we removed one participant's (Danilo) data because he chose the maximum font size (256px), so that most lines only contain one word, making it unnecessary to hesitate between lines when locating the next line. 
We had one between subject factor \textit{\textbf{Vision}} (\textit{Sighted vs. LowVision}). We validated the counterbalancing by involving another within subject factor \textbf{\textit{Order}}. We checked the normality of the different measures using Shapiro-Wilk test. 
If a measure was normally distributed, we used ANOVA for analysis and Tukey's HSD for \textit{post-hoc} comparison if significance was found. 
Otherwise, we used ART and ART contrast test for \textit{post-hoc} comparison~\cite{elkin2021aligned}. We used partial eta square ($\eta_{p}^2$) to calculate effect size for ART and ANOVA, with $0.01$, $0.06$, $0.14$ representing the thresholds of small, medium and large effects, respectively~\cite{cohen2013statistical}.}



\subsubsection{\change{Investigating the effect of visual abilities on low vision people's gaze behaviors (H3).}} We explored the effect of low vision participants' visual conditions (i.e., visual acuity, visual field) on their gaze behaviors \change{when reading in the regular mode}. The measures were the same as in Section \ref{h2}. For the same reason described in Section \ref{h2}, we removed Danilo's data in line switching analysis.
We had two between subject factors, \textit{\textbf{VisualAcuity}} and \textit{\textbf{VisualField}}. VisualAcuity had two levels---\textit{Low, High}---with \change{20/100} in the better eye as the splitting threshold~\cite{zhao2017understanding}. VisualField had two levels---\textit{\change{Limited}, Intact}---based on our visual field test. \change{We also involved a within subject factor \textbf{\textit{Order}} in our model to validate the counterbalancing. With similar analysis methods in Section \ref{h2}, we found no significant effect of Order on any of the measures.}

\subsubsection{\change{Investigating the effect of magnification modes on low vision people's gaze behaviors (H4).}} We finally investigated low vision participants' gaze behaviors under different magnification modes. \change{Since not all low vision participants used screen magnifiers in our study (e.g., some did not need magnification), we focused on those (14 participants) who used all three magnification modes.} Besides the aforementioned measures, we had another measure, \textbf{Smooth pursuit}, which refers to the slower tracking movement of the eyes to keep a moving target in the center of visual field. It happens when a low vision participant follows the lens magnifier, \change{follows a word when scrolling the text up or down in the regular mode}, and tracks a word when they pan around with the full-screen magnifier. We measured the \textbf{smooth pursuit number} \change{and \textbf{smooth pursuit duration}} to evaluate low vision participants' gaze-following behaviors.

We had two within subject factors: \textit{\textbf{MagnificationMode}} (three levels: \textit{Regular, Lens, FullScreen}) and \textbf{\textit{Order}}. We conducted analysis with the similar methods in Section \ref{h2}. However, 
\change{we found that Order had a significant effect on fixation duration (ART: $F_{(5,55)} = 2.63$, $p = 0.03$, $\eta^2_{p} = 0.19$) and smooth pursuit number (ART: $F_{(5,55)} = 3.20$, $p = 0.01$, $\eta^2_{p} = 0.23$). As Order increased, participants demonstrated shorter fixation duration and more smooth pursuits, indicating that they processed less information per fixation and tracked more frequently when moving the magnifiers. This could be explained by the fatigue accumulated along the study.}


 


\subsubsection{Qualitative analysis.} We video-recorded the initial and exit interviews and transcribed the video using an online automatic transcription service. One researcher cleaned the transcript by manually correcting the auto-transcription errors. We analyzed the data using a standard qualitative analysis method \cite{saldana2021coding}.  We developed codes using open coding. Two researchers independently coded three sample transcripts from three participants. After comparing and discussing the codes, a codebook was generated upon the two researchers' agreement. Each researcher then coded half of the remaining transcripts based on the codebook. The researchers updated the codebook upon agreement if new code emerged. We derived themes according to participants' challenges and strategies during reading.

\section{Results}
In this section, we \change{report our results regarding the four groups of hypotheses respectively, including validating gaze calibration and data quality between low vision and sighted participants (H1), comparing gaze behaviors between low vision and sighted participants (H2), and exploring the effect of visual conditions (H3) and magnification modes (H4) on low vision participants' gaze behaviors in reading.} 

\subsection{Eye Tracker Validation for Low Vision People (H1)}
We validated the gaze calibration result and assessed the data quality from the eye tracker below:

\textbf{\textit{Four-point Validation \change{(H1.1)}.}} 
We adjusted the size of calibration targets for low vision participants. Table \ref{tab:lv_task} presents the target size they selected. To validate the calibration, we compared the gaze recognition accuracy in the four-point validation between low vision and sighted participants.   
An \change{ART} analysis showed no significant difference between the gaze recognition accuracy between sighted and low vision participants \change{($F_{(1, 38)} = 0.61$, $p = 0.44$, $\eta^2_{p} = 0.02$)}.
\change{This demonstrated that the improved calibration interface for low vision participants was as effective as the typical calibration for sighted people.}  


\begin{table*}[h]
\small
\centering
\begin{tabular}{C{1.5cm}C{1.2cm}C{1.2cm}C{1.7cm}C{1.5cm}|C{1.3cm}C{1.3cm}C{1.7cm}}
\toprule
\textbf{Pseudonym}&\textbf{Target{\newline} Size}&\textbf{Acuity {\newline} Level} & \textbf{VisualField {\newline} Level} &\textbf{Dominant {\newline} Eye} &\textbf{Regular Mode}&\textbf{Lens \newline Magnifier}&\textbf{Full-screen Magnifier}\\
\midrule
Tim & 56 & Low & Intact & - & 0.39\% & 0.28\% & 2.72\%\\ 
\hline
Judy & 88 & Low & Limited  & R & 3.47\% & 1.34\% & 8.43\%\\ 
\hline
Bella & 64 & Low & Intact  & - & 4.93\% & 4.17\% & 3.73\%\\ 
\hline
Mark & 60 & \change{High} & Limited  & \change{-} & 4.56\% & 62.95\% * & 13.43\%\\ 
\hline
Kate & 36 & High & Intact  & - & 1.46\% & 1.02\% & -\\ 
\hline
Robin & 36 & High & Limited  & - & 0 & - & -\\ 
\hline
Hailey & 36 & High & Intact  & R & 0.12\% & - & -\\ 
\hline
Lucy & 56 & \change{High} & Limited  & \change{-} & 27.72\% & 1.83\% & 6.78\%\\ 
\hline
Mary & 80 & Low & Intact  & - & 1.19\% & 0.62\% & 1.28\%\\ 
\hline
Caroline & 40 & High & Intact  & L & 0.29\% & - & -\\ 
\hline
Hannah & 40 & High & Limited  & - & 0.04\% & 0.04\% & 0.38\%\\ 
\hline
Maeve & 40 & High & Intact  & L & 1.67\% & 1.81\% & 0.97\%\\
\hline
\change{May} & \change{64} & \change{High} & \change{Limited}  & \change{-} & \change{5.09\%} & \change{3.79\%} & \change{3.30\%}\\ 
\hline
\change{Piper} & \change{48} & \change{Low} & \change{Limited}  & \change{-} & \change{5.12\%} & \change{2.45\%} & \change{11.86\%}\\ 
\hline
\change{Diego} & \change{52} & \change{Low} & \change{Intact}  & \change{-} & \change{1.78\%} & \change{0.60\%} & \change{2.16\%}\\ 
\hline
\change{Danilo}& \change{92} & \change{Low} & \change{Limited}  & \change{-} & \change{16.17\%} & \change{-} & \change{11.99\%}\\ 
\hline
\change{Ryan} & \change{36} & \change{High} & \change{Intact}  & \change{-} & \change{0.19\%} & \change{0.22\%} & \change{0.77\%}\\ 
\hline
\change{Marilyn} & \change{80} & \change{Low} & \change{Limited}  & \change{L} & \change{4.61\%} & \change{3.59\%} & \change{8.68\%}\\ 
\hline
\change{Julia} & \change{52} & \change{High} & \change{Intact}  & \change{R} & \change{1.11\%} & \change{1.19\%} & \change{1.02\%}\\ 
\hline
\change{Fiona} & \change{92} & \change{Low} & \change{Limited}  & \change{R} & \change{12.57\%} & \change{12.41\%} & \change{14.88\%}\\ 

\bottomrule
\end{tabular}
\caption{Gaze calibration and collection information for the \change{20} low vision participants. \change{The table specifies each participant's calibration target size, VisualAcuity level (low vs. high with 20/100 as the threshold), VisualField level (limited vs. intact), the eye(s) we used for data collection ('L' represents left eye, 'R' represents right eye, '-' represents the mean gaze position of two eyes)}. The last three columns represent the gaze data loss rate when reading with the three magnification modes: '-' indicates they did not use the corresponding magnification; '*' indicates the corresponding piece of data was not used in our gaze pattern analysis due to high data loss rate (over 50\%).}
\label{tab:lv_task}
  \vspace{-5ex}
\end{table*}

\textbf{\textit{Data Loss \change{(H1.2)}.}} We collected participants' gaze data during reading. The data loss rate for low vision participants is shown in Table \ref{tab:lv_task}. We compared the data loss between sighted and low vision participants, and an \change{ART} analysis showed no significant difference between the two groups
\change{($F_{(1,38)} = 2.61$, $p = 0.11$, $\eta^2_{p} = 0.06$)}. \change{This indicated that the eye tracker can successfully collect similar amount of valid data from low vision people as from sighted people, verifying the feasibility of using such an eye tracker for low vision people.}  


While no statistical difference, we found that the eye tracker lost more data from low vision participants \change{($Mean = 4.62\%, SD=6.89\%$)} than from sighted participants ($Mean = 1.35\%$, \change{$SD=1.22\%$}). The data loss was commonly caused by the eye tracker's failure to recognize the user's eyes. 
Our observation revealed several reasons that may lead to the recognition failure: similar to Maus et al.'s finding~\cite{maus2020gaze}, some low vision participants had obscured pupils; some participants gradually moved their head closer than 65cm to the screen to read, exceeding the working range of the eye tracker; and some tilted their head to use their functional visual field as they usually did during reading, \change{making one eye unrecognizable}. Our results indicated that the eye tracker manufacturer should take low vision users' reading habits into consideration when developing future products.

\begin{figure*}[h]
  \includegraphics[width=\textwidth]{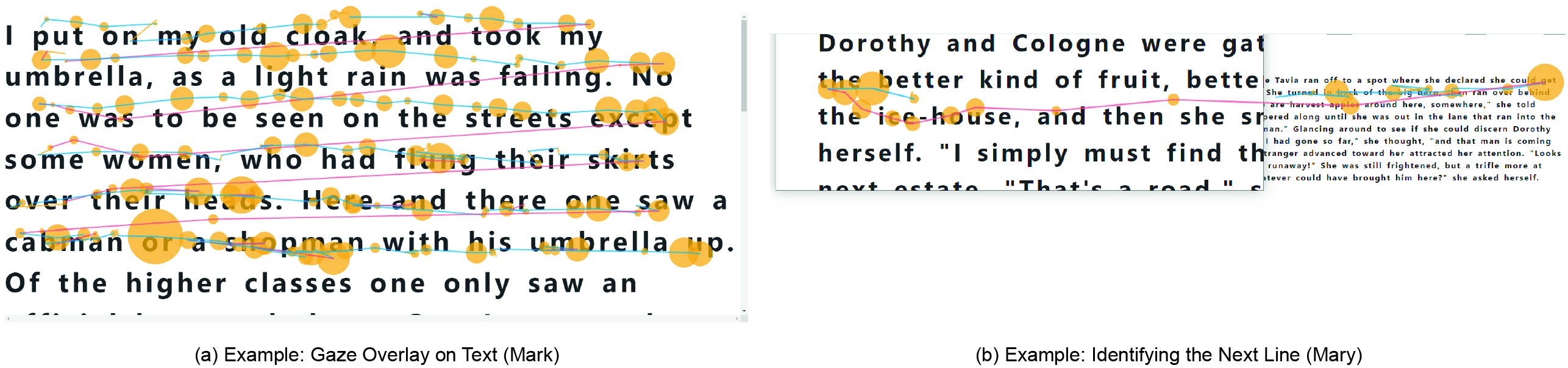}
  \caption{(a) Example of gaze overlay on the passage (Mark). Orange circles represent fixations (the bigger the longer duration); orange lines represent smooth pursuits; blue and pink lines represent forward saccades and backward saccades, respectively; (b) Example of searching for the next line under the lens magnifier (Mary): the participant hesitated between the second line and third line of the passage after finishing reading the first line.}
  \label{fig:overlay}
  \Description{
  Two images labeled (a) and (b) show examples of our low vision participants' gaze overlay on reading materials. Image (a) shows Mark's gaze overlay on the text when he read using the regular mode with large font. The position of fixations and saccades roughly align with the lines of the text. Image (b) shows Mary's gaze overlay on the text when she searched for the beginning of the next line using the lens magnifier. After finishing first line, Mary's fixations and regressive saccades first landed on the beginning of the third line in the magnification window, and then gradually navigated back to the beginning of the second line after realizing she was not on the correct line.
  }
    \vspace{-2ex}
\end{figure*}

\textbf{\textit{Gaze-Reading Alignment.}}
We also examined the alignment between the collected gaze trajectory and participants' reading progress to validate the gaze data. Fig \ref{fig:overlay}a shows an example of a low vision participant's gaze fixations and saccades overlaying on one passage read by him. \change{We extracted the timestamps  when participants read the last word of each line in a passage from audio recordings ($T_{\text{audio}}$) using VOSK~\cite{vosk} and compared $T_{\text{audio}}$ to the timestamps  of the return sweeps during line switchings from gaze data recordings ($T_{\text{gaze}}$). Using a Pearson's correlation test, we found that the time of each return sweep was highly correlated to the time of line switching during reading aloud for both sighted ($r(1013) = 1.00$, $p < 0.001$) and low vision ($r(286) = 1.00$, $p < 0.001$) participants.
}


\subsection{Comparing Low Vision and Sighted People's Gaze Behaviors \change{(H2)}}

We compared the reading and gaze behaviors of low vision and sighted people when reading in the regular mode. 
Similar to prior research~\cite{hallett2015reading, bruggeman2002psychophysics}, we found that low vision participants spent significantly longer time to complete a reading task than sighted controls 
\change{(ART: $F_{(1,78)} = 18.56$, $p < 0.001$, $\eta^2_{p} = 0.19$ )},
with low vision participants' reading time \change{($Mean=122.88s, SD=103.94s$)} being \change{1.8} times of sighted participants' reading time \change{($Mean=68.45s, SD=19.71s$)}.   
We looked into participants' detailed gaze behaviors from different aspects below.

\textbf{\textit{Fixation \change{(H2.1)}.}} An ART analysis showed that low vision participants 
had significantly more fixations than sighted participants 
\change{(ART: $F_{(1,78)} = 8.30$, $p = 0.005$, $\eta^2_{p} = 0.10$)}. However, their mean fixation duration 
was significantly shorter than sighted participants 
\change{(ART: $F_{(1,78)} = 57.8$, $p < 0.001$, $\eta^2_{p} = 0.43$}). This indicated that low vision people's gaze tended to fixate more frequently during active reading, but each fixation was shorter than sighted people, representing less amount of information being processed \change{per fixation} and \change{lower reading efficiency by low vision people}~\cite{moffitt1980evaluation}. 

\textbf{\textit{Sacaade \change{(H2.2)}.}} Low vision participants conducted significantly more \change{(ART: $F_{(1,78)} = 4.54$, $p = 0.04$, $\eta^2_{p} = 0.05$)} but shorter (\change{ANOVA: $F_{(1,78)} = 36.74$, $p < 0.001$, $\eta^2_{p} = 0.32$}) forward saccades 
than sighted participants. 
This indicated that the perceptual span of low vision people---the number of letters taken in during a single glance---was shorter than sighted people~\cite{rubin2009role}. They thus needed to fixate and skim more times to read the same content.   
Meanwhile, we found that low vision participants demonstrated significantly more regressive saccades 
than sighted participants 
\change{(ART: $F_{(1,78)} = 5.50$, $p = 0.02$, $\eta^2_{p} = 0.07$)}. However, no significant difference was found in terms of revisitation rate \change{(ART: $F_{(1,78)} = 1.48$, $p = 0.23$, $\eta^2_{p} = 0.02$)}, showing that low vision people may be able to scan the text to locate the next fixation point (i.e., follow a line) as accurately as sighted people.  

\textbf{\textit{Lines Switching \change{(H2.3)}.}} Locating the next line can be a challenge for low vision people~\cite{verghese2021eye}. \change{10 out of 20} low vision participants reported they had difficulty locating the next line. We found that low vision participants searched significantly more lines 
than sighted participants 
when switching lines (\change{ART: $F_{(1,76)} = 11.64$, $p = 0.001$, $\eta^2_{p} = 0.13$}). 

\begin{figure*}[h]
  \includegraphics[width=0.8\textwidth]{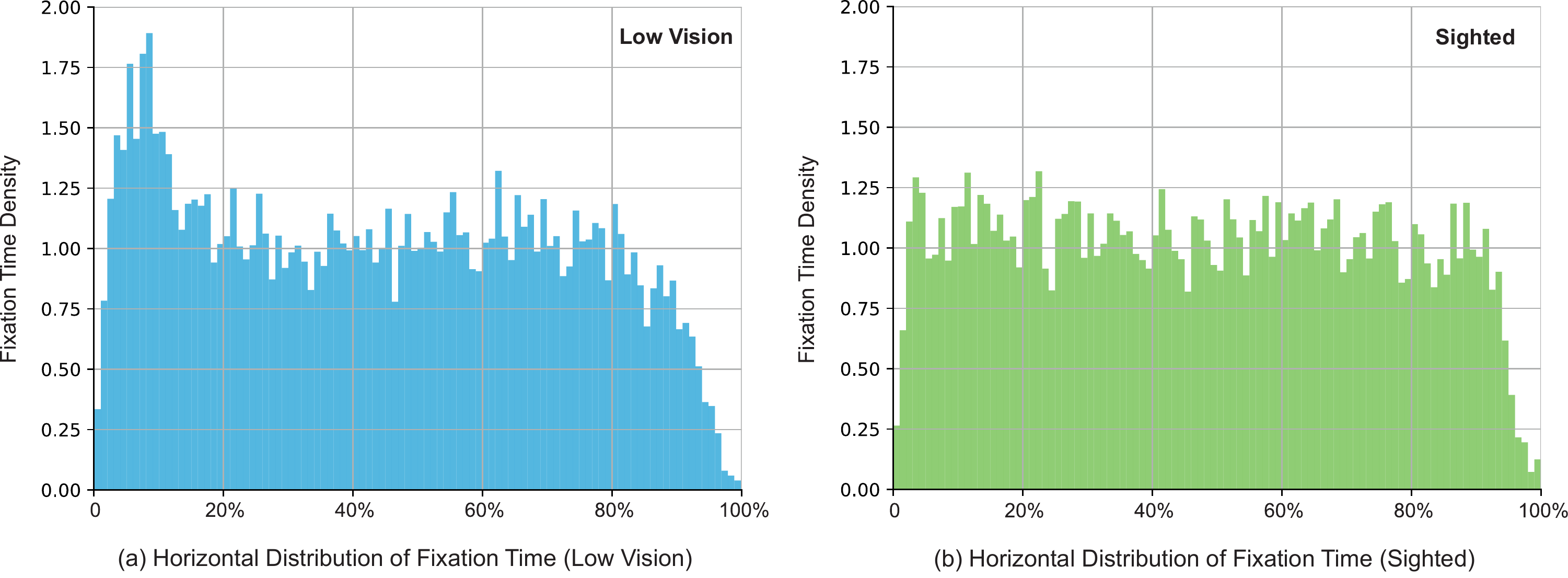}
  \caption{\change{(a) The distribution of fixation time along the horizontal axis of the reading interface for low vision participants (mean fixation time across all participants and all passages);  (b) The distribution of fixation time along the horizontal axis of the reading interface for sighted participants.}}
  \label{fig:hist_fix}
  \Description{
  Two figures (a) and (b) show the distributions of fixation time along the horizontal axis of the reading interface for low vision and sighted participants. Figure (a) shows the fixation time density of low vision participants on the Y axis against the horizontal percentage of the text from 0 to 100\% on the X axis. There is a noticeable spike of the fixation time density within the first 10\% of the text. Figure (b) shows the fixation time density of sighted participants on the Y axis against the horizontal percentage of the text from 0 to 100\% on the X axis. The fixation time density is consistent across the horizontal axis.
  }
    \vspace{-2ex}
\end{figure*}

We further examined how low vision people located a line. \change{We investigated the distribution of participants' fixation time (i.e., the total time of all fixations happened at a particular area) along the horizontal axis of the reading interface (Fig. \ref{fig:hist_fix})}, and found that low vision participants spent significantly longer fixation time at the first 10\% 
of each line than sighted people (\change{ART: $F_{(1,76)} = 26.93$, $p < 0.001$, $\eta^2_{p} = 0.26$}). Specifically, low vision participants spent \change{14.87\% ($SD=4.30\%$)} of the total reading time fixating at the first 10\% of a line, while sighted participants used 
\change{11.27\% ($SD=2.34\%$)} of the total reading time. The result suggested that low vision participants spent more time locating and confirming the correct line when switching lines. According to low vision participants, a common strategy of identifying the next line was checking if the content of the new line made sense based on the context of the passage. For example, 
\change{Judy always remembered the last several words of a line during reading, so that she could compare the beginning of the next lines to see which line made the most sense.} 
This explained the higher cognitive processing load (longer fixation time) observed from low vision participants at the beginning of each line during line switching.

\subsection{Effects of Different Visual Conditions on Gaze Patterns \change{(H3)}}
We studied the effect of different low vision conditions on low vision participants' gaze behaviors.

\textbf{\textit{Fixation \change{(H3.1)}.}} Our result showed that participants with low visual acuity demonstrated significantly 
more fixations (\change{ART: $F_{(1,36)} = 5.18$, $p = 0.03$, $\eta^2_{p} = 0.13$}) but shorter fixation duration (\change{ART: $F_{(1,36)} = 6.01$, $p = 0.02$, $\eta^2_{p} = 0.14$}) than those with relatively high visual acuity. \change{Similarly}, participants with limited visual field had more fixations (\change{ART: $F_{(1,36)} = 4.69$, $p = 0.04$, $\eta^2_{p} = 0.12$}) \change{ and shorter fixation duration (ART: $F_{(1,36)} = 7.88$, $p = 0.01$, $\eta^2_{p} = 0.18$)} than those with intact visual field. 
An interaction between visual field and visual acuity was found. \change{Through a \textit{post-hoc} contrast test for ART}, we found participants with visual field loss and low visual acuity had significantly shorter fixation duration than participants with low visual acuity only 
\change{($t_{(36)} = 3.08$, $p = 0.02$)},
participants with visual field loss only \change{($t_{(36)}  = 2.88$, $p = 0.03$)},
and participant with both high visual acuity and intact visual field 
\change{($t_{(36)} = 3.24$, $p = 0.01$)}.
Our result suggested that \change{both low visual acuity and visual field loss lead to} reduced amount of information low vision people can perceive \change{per fixation}, and \change{experiencing visual field loss and low visual acuity at the same time} can further impact low vision people's ability to access information.

\textbf{\textit{Saccade \change{(H3.2)}.}} In terms of saccade, we found no significant effect of visual acuity \change{(ART: $F_{(1,36)} = 2.14$, $p = 0.15$, $\eta^2_{p} = 0.06$)} or \change{visual field (ART: $F_{(1,36)} = 1.12$, $p = 0.30$, $\eta^2_{p} = 0.03$)} on the number of forward saccades from low vision participants. However, we found that participants with low acuity had significantly shorter forward saccade length \change{than participants with high visual acuity (ANOVA: $F_{(1,36)} = 12.00$, $p = 0.001$, $\eta^2_{p} = 0.25$). Similarly, we found participants with limited visual field had significantly shorter forward saccade length than those with intact visual field (ANOVA: $F_{(1,36)} = 7.97$, $p = 0.01$, $\eta^2_{p} = 0.18$).} This implied that \change{low visual acuity and} limited visual field can \change{both} affect low vision participants' length of perceptual span. 
\change{For example, as Danilo explained, he could only focus on a small area at a time when reading due to his limited visual field, and since his visual acuity was low, he had to recognize letter by letter, because according to him, the letters were "mashed together".}

\change{However, we found no significant effect of either visual acuity or visual field on the number of regressive saccades (visual acuity, ART: $F_{(1,36)} = 3.35$, $p = 0.08$, $\eta^2_{p} = 0.09$; visual field, ART: $F_{(1,36)} = 2.71$, $p = 0.11$, $\eta^2_{p} = 0.07$), and revisitation rate (visual acuity, ART: $F_{(1,36)} = 1.85$, $p = 0.18$, $\eta^2_{p} = 0.05$; visual field, ART: $F_{(1,36)} = 1.83$, $p = 0.18$, $\eta^2_{p} = 0.05$).}



\textbf{\textit{Lines Switching \change{(H3.3)}.}} We looked into how low vision participants with different visual abilities switched lines, and found no significant effect of visual field \change{(ART: $F_{(1,34)} = 0.0002$, $p = 0.99$, $\eta^2_{p} < 0.001$)} on the number of searched lines during line switching. \change{However, there was a trend towards a significant effect of visual acuity \change{(ART: $F_{(1,34)} = 3.62$, $p = 0.07$, $\eta^2_{p} = 0.10$)} on the number of searched lines. More research is needed to further investigate the impact of visual acuity on low vision people's line switching behaviors.} 


\subsection{Gaze Behaviors under Different Magnification Modes \change{(H4)}} 
 We compared the reading performance and gaze patterns of low vision participants when reading with different magnification modes. To start off, it was not surprising that magnification modes had significant effect on participants' reading time \change{(ART: $F_{(2,26)} = 16.45$, $p < 0.001$, $\eta^2_{p} = 0.56$)}. A \textit{post-hoc} contrast test for ART showed that, participants read much slower using the lens magnifier \change{($t_{(26)} = -5.31$, $p < 0.001$)} or full-screen magnifier \change{($t_{(26)} = -4.53$, $p < 0.001$)} than using the regular mode with increased font size. 
 \change{Nine} participants mentioned that the \change{hand-eye} coordination required by screen magnifiers made reading more difficult. 
 One participant (Bella) told us that she read slower on purpose to make the lens magnifier move stably. No significant difference in reading time was found between the lens magnifier and full-screen magnifier \change{($t_{(26)} = 0.79$, $p = 0.71$)}. 
 
 \textbf{\textit{Fixation \change{(H4.1)}.}} Although there was no significant effect of magnification modes on the number of fixations \change{(ART: $F_{(2,26)} = 0.03$, $p = 0.97$, $\eta^2_{p} = 0.003$)}, we found a significant effect of magnification modes on the mean fixation duration \change{(ART: $F_{(2,55)} = 38.60$, $p < 0.001$, $\eta^2_{p} = 0.58$)}. A \textit{post-hoc} contrast test showed that the lens magnifier and full-screen magnifier both led to significantly shorter fixation duration than the regular mode with increased font size \change{(lens: $t_{(55)} = 8.61$, $p < 0.001$; full-screen: $t_{(55)} = 5.78$, $p < 0.001$). Moreover, participants showed significantly shorter fixation duration when using the lens magnifier than using the full-screen magnifier ($t_{(55)} = -2.84$, $p = 0.02$)}. One possible explanation was that the dynamic changing of text layout (due to the lens movement and the panning) made it more difficult for low vision participants to focus and fixate on individual words, leading to shorter fixation duration. 
 \change{Eight} participants 
 complained that the control of the magnifiers made it harder to keep track of their position, \change{especially for the lens magnifier. Due to the high magnification level, a small deviation on the position of the magnification window would cause a whole line to disappear. 
 As Diego commented, \textit{``I drop [the magnification window] down [a little] too far like that, then it can throw me off from one end [of the window] to the other.''}}


\textbf{\textit{Saccades \change{(H4.2)}.}} There was no significant effect of magnification modes on the number of forward saccades (\change{ART: $F_{(2,26)} = 3.07$, $p = 0.06$, $\eta^2_{p} = 0.19$}) and mean forward saccade length (\change{ART: $F_{(2,26)} = 1.34$, $p = 0.28$, $\eta^2_{p} = 0.09$}). \change{However, a significant effect was found on the number of regressive saccades (ART: $F_{(2,26)} = 6.19$, $p = 0.01$, $\eta^2_{p} = 0.32$). Through a \textit{post-hoc} contrast test, we found that low vision participants had more regressive saccades when using the lens magnifier \change{($t_{(26)} = -2.85$, $p = 0.02$)} and full-screen magnifier \change{($t_{(26)} = -3.22$, $p = 0.01$)} than using the regular mode.} 
Furthermore, we found magnification modes had a significant effect on revisitation rate \change{(ART: $F_{(2,26)} = 6.62$, $p = 0.005$, $\eta^2_{p} = 0.34$)}. Using a \change{\textit{post-hoc} contrast test}, we found participants showed higher revisitation rate using the lens magnifier \change{($t_{(26)} = -3.19$, $p = 0.01$)} and the full-screen magnifier \change{($t_{(26)} = -3.12$, $p = 0.01$)} than using the regular mode with increased font size. No significant difference in the number of regressive saccades \change{($t_{(26)} = -0.37$, $p = 0.93$)} and revisitation rate (\change{$t_{(26)} = 0.06$, $p = 1.00$}) was found between the lens and full-screen magnifier.
This could be explained by participants' difficulty of keeping track of their reading positions when using screen magnifiers (Hannah, Judy, Bella, Lucy, Mark and \change{May}). One strategy reported by participants was to go back and check if they had read the content to locate where they were reading, \change{thus leading to more regressive saccades and revisitations}.

\change{For the lens magnifier, we observed that \change{12} low vision participants requested to increase the width of the magnification window.} We thus investigated how the size of the magnification window affected participants' gaze behaviors. We normalized the window width by calculating the number of letters that can be horizontally covered by the window. We then examined the relationship between the mean forward saccade length and the normalized window width. Using a \change{Pearson's correlation} test, we found a positive correlation between the two variables (\change{$r(26) = 0.72$, $p < 0.001$}). We further found a negative correlation between reading time and the normalized window width \change{($r(26) = -0.67$, 
$p < 0.001$), which echoed the result in Legge et al.~\cite{legge1985psychophysics}.} Combined with the prior finding that reading speed was correlated with forward saccade length~\cite{rubin2009role}, our result suggested that a wider magnification window can potentially benefit reading speed by improving perceptual span.




\textbf{\textit{Lines Switching \change{(H4.3)}.}} We found that the effect of magnification modes on the number of searched lines during line switching was significant \change{(ART: $F_{(2,26)} = 5.70$, $p = 0.01$, $\eta^2_{p} = 0.30$)}. A \textit{post-hoc} contrast test showed that participants searched significantly more lines to locate the next line when using the lens magnifier than using the regular mode \change{($t_{(26)} = -3.35$, $p = 0.01$)}. 
\change{Fig \ref{fig:overlay}b shows an example where the participant hesitated between the second and third line before landing on the correct line using the lens magnifier.} \change{Six} participants (Judy, Mark, Tim, \change{May, Marilyn and Diego}) explicitly mentioned that the small magnification window in the lens magnifier caused the challenge of locating the next line. Judy further explained that she could easily move the magnifier out of the text area to the white margin on the left because of limited context information presented in the magnification window. To successfully locate the next line, \change{five participants (Judy, Mark, Marilyn, Diego and Julia)} carefully moved the lens magnifier one line down and then back tracked to find the correct location. 

Moreover, for the lens magnifier, we found a negative correlation between the number of searched lines and the normalized height of the window (i.e., the number of lines covered by the window at a time) with Pearson's correlation test \change{($r(26) = -0.60$, $p < 0.001$)}. This meant that the more lines that can be viewed within the magnification window, the fewer lines participants needed to search to locate the correct lines. However, not all participants preferred a taller magnification window. Three participants (Bella, Mary and Lucy) preferred a shorter window since too much content around the current line can be distracting. As Mary commented, \textit{``In some ways [a shorter lens magnifier] was easier because you didn't have so many possibilities [of lines]. Your options were limited to what you needed to choose.''}

When asked about potential improvements on the magnification tools, four participants (Hannah, Judy, Lucy and Robin) suggested highlighting the next line and labeling the index of each line at the beginning to assist with line switching. To reduce the effect of `panning around' on line switching, \change{seven} participants suggested using a larger screen to present all enlarged text, or breaking the text into pages and allowing page flipping (Bella and Tim). As Tim commented, \textit{``The more I have to [move the screen magnifier], the more I'm going to lose my place.''} 
Furthermore, two participants (Lucy and Tim) suggested marking where their eyes stopped if they look away from the screen to prevent them from losing track of where they were. 

\textbf{\textit{Smooth Pursuits.}} Unlike reading static text, when moving the text by either scrolling, panning, or moving the magnification window, smooth pursuits happened. Our analysis showed that magnification modes had a significant effect on the number \change{(ART: $F_{(2,55)} = 88.08$, $p < 0.001$, $\eta^2_{p} = 0.76$)} and mean duration \change{(ANOVA: $F_{(2,26)} = 24.38$, $p < 0.001$, $\eta^2_{p} = 0.65$)} of smooth pursuits.
\change{\textit{Post-hoc} comparisons showed that} when low vision participants used the lens magnifier or the full-screen magnifier, the number of smooth pursuits was significantly higher than the regular mode \change{(lens: $t_{(55)} = -12.50$, $p < 0.001$; full-screen: $t_{(55)} = -9.98$, $p < 0.001$)}. Moreover, \change{participants conducted more smooth pursuits when using the lens magnifier than the full-screen magnifier ($t_{(55)} = 2.53$, $p = 0.04$).} 
The mean duration of smooth pursuits when using the two screen magnifiers was also significantly longer than using the regular mode \change{(lens: $t_{(26)} = -6.58$, $p < 0.001$; full-screen: $t_{(26)}= -5.32$, $p < 0.001$)}.
However, there was no significant difference in the duration of smooth pursuits between the two magnifiers  (\change{$t_{(26)} = 1.26$, $p = 0.43$}). Due to the higher number and longer duration of smooth pursuits, participants (Mary and Hannah) felt more tired when using the \change{lens} magnifier than \change{the other two magnification modes} since they had to constantly track the position of the text in the moving window. As Hannah complained, \textit{``The window is always moving, searching for the right position makes my eyes tired.''}

\section{Discussion}
\change{We contributed the first research that (1) investigated the suitable gaze calibration and collection for low vision people via commercial eye trackers and (2) thoroughly explored low vision people's unique gaze behaviors and challenges during reading. 
We answer our four research questions by responding to our four groups of hypotheses.}

\change{With the adjustable calibration interface and dominant-eye-based data collection, we found that commercial eye trackers (e.g., Tobii eye tracker) can achieve comparable quality of gaze data between low vision and sighted people (H1), with no significant difference in the gaze recognition accuracy (H1.1) and data loss (H1.2). The high alignment between participants' gaze trajectory and their reading progress also validated low vision participants' data quality. These results highlighted the potential of commercial eye trackers in low vision research. With commercial eye trackers increasingly integrated in everyday devices, our research opened up a new research direction for gaze-based vision enhancement technology for low vision users.}

\change{We further characterized low vision people's gaze challenges in reading via their gaze patterns. We found that low vision people demonstrated different gaze behaviors from sighted people (H2), with more but shorter fixations (H2.1) as well as more but shorter forward saccades (H2.2), indicating their difficulty with information perception per fixation and skimming. We also identified low vision people's challenges with line switching due to their significant higher number of searched lines (H2.3). Moreover, we found that different visual abilities (H3) and magnification modes (H4) can affect low vision people's gaze behaviors differently. Specifically, both low visual acuity and visual field loss led to more but shorter fixations (H3.1) and shorter forward saccades (H3.2). However, no significant effect of visual abilities (only a trend in visual acuity) was found on line switching behaviors (H3.3). In terms of magnification modes, we found the regular mode with increased font size more effective than the lens and full-screen magnifier, with longer fixation duration (H4.1), fewer regressive saccades, lower revisitation rates (H4.2), and fewer searched lines during line switching (H4.3).}   
Unlike prior work that infers low vision people's visual experiences and challenges via reading performances 
~\cite{bruggeman2002psychophysics, hallett2017screen,
harland1998psychophysics, moreno2021exploratory, ahn1995psychophysics}, our work draw direct evidence on participants' gaze data at the word and sentence level to develop a deep understanding of their challenges. This fine-grained investigation allows us to unfold low vision people's reading difficulties (e.g., locating next line correctly) and derive design implications for more targeted vision enhancement technology based on low vision users' detailed gaze behaviors. 
We discuss our results by identifying the limitations and potential improvements for current eye tracking technology, as well as deriving implications to guide the design of gaze-based technology for low vision. 



\subsection{Accessible Eye Tracking Technology for Low Vision Users}
\subsubsection{\change{Accessible Calibration Interface for Different Visual Conditions}}
\change{The conventional calibration interface (e.g., Tobii Pro Lab) did not consider low vision people's visual abilities and eye characteristics.} While we refined the calibration interface by customizing the target size for low vision people, addressing the invisible target issue, \change{our calibration process could still be challenging for some low vision participants. For example, some participants could not focus on a target for a long duration required by the calibration due to their visual conditions, such as Nystagmus (Caroline) and dry eyes (Fiona). To address this, future calibration could consider data collection on more targets with shorter duration for each target.} Since different low vision users may have different visual abilities and thus different needs for the calibration process, adaptations based on the user's visual conditions can be integrated, automatically adjusting the calibration interface, such as the number, color, and size of the calibration targets, and the duration of each target.


\subsubsection{\change{Data Collection by Considering the Dominant Eye.}} 
 \change{Binocular data collection strategy can cause low accuracy if the user demonstrates inconsistent eye movement or irregular pupil appearance \cite{maus2020gaze}. We addressed this via dominant-eye-based data collection if a low vision user had an obvious dominant eye (i.e., the weaker eye does not follow the dominant eye or one eye is not recognizable). Our data collection strategy resulted in relatively high accuracy that is comparable to sighted people. Not only for low vision users, prior work with sighted participants also shows that the dominant eye can fixate at a target more accurately and precisely than the non-dominant eye~\cite{simonsz1991covering, gibaldi2017evaluation}. As such, the dominant eye should be considered in data collection when high accuracy is required. While some low-tech tests (e.g., hole-in-the-card test~\cite{durand1910method}) are usually needed to determine eye dominance without prior knowledge, future eye tracking technology should consider how to automatically detect the dominant eye. Despite the advantages of dominant-eye-based data collection, this strategy can also be volatile when occlusion of the dominant eye occurs due to head and body movement. Tradeoff should be considered to suitably adopt binocular or dominant-eye-based data collection.}
 

\subsubsection{\change{Requirements on the User Position}}
Current eye trackers impose strict requirements on the user's sitting position, including the distance between the user's eyes and the screen and the relative height of the user's head to the screen. \change{However, this could conflict with low vision people's common reading habits. For example, many low vision people tend to get very close to the screen to read \cite{maus2020gaze, szpiro2016people} but exceeded the standard distance range required by the eye tracker.} \change{Future eye tracker design should consider the unique reading habits of low vision people.} Moreover, wearable eye trackers could be considered to reduce the burdens on low vision users to fulfil the requirements of screen-based eye trackers. 

\subsubsection{\change{Transparency of the Data Collection Status.}} \change{Beyond the essential issues with current gaze calibration and collection methods, transparency of the data collection status should also be improved to enhance the eye tracker usability.} For example, some participants (Piper) felt confused about when the data collection for a target started and ended due to the lack of feedback for data collection status. Moreover, in our study, we found it hard to determine and maintain the data collection quality \change{in real time, which led to unusable data and discarded participant samples}. \change{It is thus important to provide proper feedback on an eye tracking interface to indicate the data collection status (e.g., beginning, end, and failure of data collection), and prompt instructions to the user timely when necessary to ensure high quality data collection.} For example, when the user's eyes become undetectable, the system should prompt the user to adjust their position in an explicit \change{and accessible} way.

\subsection{Design Implications for Gaze-Based Assistive Technology}
Our work highlights the potential of detecting low vision users' eye movement events using commercial eye trackers. We propose the following design implications for future assistive technologies using eye tracking.

\subsubsection{Real-Time Support for Line Following and Line Switching}
Locating the next line and keeping track of the current line is a big challenge for low vision people, especially when using screen magnifiers. With eye tracking, we can identify the line a user is reading and their intent to switch lines by recognizing the regressive saccade (i.e., return sweep) at the end of each line. As such, visual augmentations could be designed to provide real-time support, such as highlighting the current or next line~\cite{gowases2011text}, or dynamically adjusting the line spacing based on the users' gaze position. Participants mentioned different strategies to orient themselves in a reading material, for example, remembering the last few words of the line to see if the next line made sense (Judy, Lucy, \change{Danilo, Marilyn and Julia}), or memorizing the first several words of the prior line to see if they had read the same line (Lucy, Tim, \change{Piper and Fiona}). With eye tracking, an assistive system can be designed to remind the user of those key words at the right timing to help them locate the correct line. Moreover, we can detect irregular gaze behaviors during reading, such as losing track of a line due to looking away from the screen or the moving of the magnifier, and generate feedback to notify the user where they were reading to improve their reading efficiency.

\subsubsection{\change{Support for Unrecognizable Words}}
While many low vision people prefer using vision to access information~\cite{szpiro2016people, zhao2015foresee}, some participants  had trouble recognizing words due to low visual acuity or visual field loss, such as distorted words (Judy) and missing letters (Mary, Bella and \change{Julia}). With eye tracking technique, we can potentially recognize these issues and provide corresponding assistance. For example, when a long fixation at a word or frequent revisitations are detected, the system can directly read the nearby words or phrases to the user to improve their reading experience. 



\subsubsection{Hands-Free and Context-Aware Screen Magnifier}
Eye tracking technique has the potential to improve screen magnifiers. Moving and tracking the screen magnifier can diminish reading performance and experience \cite{moreno2021exploratory, ahn1995psychophysics, szpiro2016people, hallett2015reading, hallett2017screen}. \change{Nine} low vision participants in our study echoed this problem of hand-eye coordination. \change{Thus, instead of moving the magnifier around with a mouse, hands-free screen magnifiers controlled by gaze could be more desirable~\cite{maus2020gaze, aguilar2017evaluation}.}
\change{Another drawback of screen magnifiers is fixed local-view to global-view ratio.}
Our study revealed that different window size for the lens magnifier \change{had different pros and cons:} taller windows (i.e., window containing more lines) contain more contextual information, \change{but can be more distracting}; while shorter windows reduce distraction, \change{but lacks sufficient context}. Future technology can consider a context-aware magnifier, which detects the user's needs and adjust the magnification window accordingly.
For example, when the system detects that the user loses track or is switching lines, the window size would be automatically enlarged to contain more contextual information; when the user is actively reading a line, the window size would be reduced to include fewer lines to reduce distraction. 

\subsection{Limitations \& Future Work}
Our research has limitations.  
\change{First, our low vision participants presented a variety of visual conditions, which can potentially reduce the statistical power. However, low vision is complex and it is very difficult to recruit many low vision participants with the same visual abilities. For example, participants who have similar visual acuity could have very different field of view; participants with the same diagnosed condition could experience different severity.
As such, as the first step of exploration, we broadly recruited participants with different visual abilities and analyzed the effect of visual abilities on low vision people's gaze behaviors. This study helped us identify particular low vision conditions that worth further investigation, for example, people with Nystagmus are particularly hard for eye tracker to track their gaze. In the future, we will focus on particular low vision groups and recruit more participants for a more thorough analysis. Second, our low vision group and sighted group had a big age difference. While prior research has shown that age does not affect low vision people's gaze behaviors during reading~\cite{bowers2001eye}, future research should consider recruiting a sighted control group at the similar age level to the low vision group to remove the potential effect.}
Lastly, to support the analysis of gaze data, we asked participants to read the passages aloud to confirm their reading progress. However, it is unclear whether our findings could be generalized to silent reading, given that reading aloud is slower and involves more frequent fixations than silent reading~\cite{rayner200935th}. 
Future research should examine low vision people's gaze behaviors in silent reading to better simulate their daily reading experience.

\section{Conclusion}
\change{In this paper, we explored the potential of using a commercial eye tracker for low vision people and investigated their reading gaze behaviors by conducting a study with 20 low vision participants and 20 sighted controls. We improved the traditional calibration interface and validated the effectiveness and data quality of the eye tracker for low vision users. We further explored low vision people's gaze behaviors, revealing their unique gaze patterns compared to sighted people, as well as the effect of different visual abilities and magnification modes on low vision people's gaze behaviors. 
We hope that our work will inspire the design and development of gaze-based assistive technology for low vision people.}

\begin{acks}
We would like to thank Davit Khachatryan for his contribution to participant recruitment and data analysis, as well as our study participants for their valuable feedback. This work was partially supported by the University of Wisconsin--Madison Office of the Vice Chancellor for Research and Graduate Education with funding from the Wisconsin Alumni Research Foundation, and the Expanding Our Vision 2021 Award from the McPherson Eye Research Institute at the University of Wisconsin--Madison. 
\end{acks}


\bibliographystyle{ACM-Reference-Format}
\bibliography{sample-base}


\begin{thebibliography}{78}


\ifx \showCODEN    \undefined \def \showCODEN     #1{\unskip}     \fi
\ifx \showDOI      \undefined \def \showDOI       #1{#1}\fi
\ifx \showISBNx    \undefined \def \showISBNx     #1{\unskip}     \fi
\ifx \showISBNxiii \undefined \def \showISBNxiii  #1{\unskip}     \fi
\ifx \showISSN     \undefined \def \showISSN      #1{\unskip}     \fi
\ifx \showLCCN     \undefined \def \showLCCN      #1{\unskip}     \fi
\ifx \shownote     \undefined \def \shownote      #1{#1}          \fi
\ifx \showarticletitle \undefined \def \showarticletitle #1{#1}   \fi
\ifx \showURL      \undefined \def \showURL       {\relax}        \fi
\providecommand\bibfield[2]{#2}
\providecommand\bibinfo[2]{#2}
\providecommand\natexlab[1]{#1}
\providecommand\showeprint[2][]{arXiv:#2}

\bibitem[Aguilar and Castet(2017)]%
        {aguilar2017evaluation}
\bibfield{author}{\bibinfo{person}{Carlos Aguilar} {and} \bibinfo{person}{Eric
  Castet}.} \bibinfo{year}{2017}\natexlab{}.
\newblock \showarticletitle{Evaluation of a gaze-controlled vision enhancement
  system for reading in visually impaired people}.
\newblock \bibinfo{journal}{\emph{PLoS One}} \bibinfo{volume}{12},
  \bibinfo{number}{4} (\bibinfo{year}{2017}), \bibinfo{pages}{e0174910}.
\newblock


\bibitem[Ahn and Ledge(1995)]%
        {ahn1995psychophysics}
\bibfield{author}{\bibinfo{person}{Sonia~J Ahn} {and} \bibinfo{person}{Gordon~E
  Ledge}.} \bibinfo{year}{1995}\natexlab{}.
\newblock \showarticletitle{Psychophysics of reading—XIII. Predictors of
  magnifier-aided reading speed in low vision}.
\newblock \bibinfo{journal}{\emph{Vision Research}} \bibinfo{volume}{35},
  \bibinfo{number}{13} (\bibinfo{year}{1995}), \bibinfo{pages}{1931--1938}.
\newblock


\bibitem[Apple(2022)]%
        {maczoom}
\bibfield{author}{\bibinfo{person}{Apple}.} \bibinfo{year}{2022}\natexlab{}.
\newblock \bibinfo{title}{How to zoom in or out on Mac}.
\newblock
\newblock
\newblock
\shownote{Available online at: \url{https://support.apple.com/en-us/HT210978},
  last accessed on 8/23/2022}.


\bibitem[Asada(2016)]%
        {biggerbrighter}
\bibfield{author}{\bibinfo{person}{Kazunori Asada}.}
  \bibinfo{year}{2016}\natexlab{}.
\newblock \bibinfo{title}{Brighter and Bigger}.
\newblock
\newblock
\newblock
\shownote{Available online at:
  \url{http://asada.website/brighterandbigger/e/index.html}, last accessed on
  12/06/2022}.


\bibitem[Bax(2013)]%
        {bax2013cognitive}
\bibfield{author}{\bibinfo{person}{Stephen Bax}.}
  \bibinfo{year}{2013}\natexlab{}.
\newblock \showarticletitle{The cognitive processing of candidates during
  reading tests: Evidence from eye-tracking}.
\newblock \bibinfo{journal}{\emph{Language Testing}} \bibinfo{volume}{30},
  \bibinfo{number}{4} (\bibinfo{year}{2013}), \bibinfo{pages}{441--465}.
\newblock


\bibitem[Bowers et~al\mbox{.}(2001)]%
        {bowers2001eye}
\bibfield{author}{\bibinfo{person}{Alex~R Bowers}, \bibinfo{person}{Jan~E
  Lovie-Kitchin}, \bibinfo{person}{Russell~L Woods}, {et~al\mbox{.}}}
  \bibinfo{year}{2001}\natexlab{}.
\newblock \showarticletitle{Eye movements and reading with large print and
  optical magnifiers in macular disease}.
\newblock \bibinfo{journal}{\emph{Optometry and Vision Science}}
  \bibinfo{volume}{78}, \bibinfo{number}{5} (\bibinfo{year}{2001}),
  \bibinfo{pages}{325--334}.
\newblock


\bibitem[Bowers and Poletti(2017)]%
        {bowers2017microsaccades}
\bibfield{author}{\bibinfo{person}{Norick~R Bowers} {and}
  \bibinfo{person}{Martina Poletti}.} \bibinfo{year}{2017}\natexlab{}.
\newblock \showarticletitle{Microsaccades during reading}.
\newblock \bibinfo{journal}{\emph{PloS one}} \bibinfo{volume}{12},
  \bibinfo{number}{9} (\bibinfo{year}{2017}), \bibinfo{pages}{e0185180}.
\newblock


\bibitem[Bradley(1958)]%
        {bradley1958complete}
\bibfield{author}{\bibinfo{person}{James~V Bradley}.}
  \bibinfo{year}{1958}\natexlab{}.
\newblock \showarticletitle{Complete counterbalancing of immediate sequential
  effects in a Latin square design}.
\newblock \bibinfo{journal}{\emph{J. Amer. Statist. Assoc.}}
  \bibinfo{volume}{53}, \bibinfo{number}{282} (\bibinfo{year}{1958}),
  \bibinfo{pages}{525--528}.
\newblock


\bibitem[Brown et~al\mbox{.}(1977)]%
        {brown1977clinically}
\bibfield{author}{\bibinfo{person}{Brian Brown}, \bibinfo{person}{Anthony~J
  Adams}, \bibinfo{person}{Arthur Jampolsky}, {and} \bibinfo{person}{Michael
  Muegge}.} \bibinfo{year}{1977}\natexlab{}.
\newblock \showarticletitle{A clinically useful eye movement recording system.}
\newblock \bibinfo{journal}{\emph{American Journal of Optometry and
  Physiological Optics}} \bibinfo{volume}{54}, \bibinfo{number}{1}
  (\bibinfo{year}{1977}), \bibinfo{pages}{56--60}.
\newblock


\bibitem[Bruggeman and Legge(2002)]%
        {bruggeman2002psychophysics}
\bibfield{author}{\bibinfo{person}{Hugo Bruggeman} {and}
  \bibinfo{person}{Gordon~E Legge}.} \bibinfo{year}{2002}\natexlab{}.
\newblock \showarticletitle{Psychophysics of reading. XIX. Hypertext search and
  retrieval with low vision}.
\newblock \bibinfo{journal}{\emph{Proc. IEEE}} \bibinfo{volume}{90},
  \bibinfo{number}{1} (\bibinfo{year}{2002}), \bibinfo{pages}{94--103}.
\newblock


\bibitem[Bullimore and Bailey(1995)]%
        {bullimore1995reading}
\bibfield{author}{\bibinfo{person}{Mark~A Bullimore} {and}
  \bibinfo{person}{Ian~L Bailey}.} \bibinfo{year}{1995}\natexlab{}.
\newblock \showarticletitle{Reading and eye movements in age-related
  maculopathy}.
\newblock \bibinfo{journal}{\emph{Optometry and Vision Science}}
  \bibinfo{volume}{72}, \bibinfo{number}{2} (\bibinfo{year}{1995}),
  \bibinfo{pages}{125--138}.
\newblock


\bibitem[Cephei(2022)]%
        {vosk}
\bibfield{author}{\bibinfo{person}{Alpha Cephei}.}
  \bibinfo{year}{2022}\natexlab{}.
\newblock \bibinfo{title}{VOSK Offline Speech Recognition API}.
\newblock
\newblock
\newblock
\shownote{Available online at: \url{https://alphacephei.com/vosk/}, last
  accessed on 12/11/2022}.


\bibitem[Cheng et~al\mbox{.}(2015)]%
        {cheng2015gaze}
\bibfield{author}{\bibinfo{person}{Shiwei Cheng}, \bibinfo{person}{Zhiqiang
  Sun}, \bibinfo{person}{Lingyun Sun}, \bibinfo{person}{Kirsten Yee}, {and}
  \bibinfo{person}{Anind~K Dey}.} \bibinfo{year}{2015}\natexlab{}.
\newblock \showarticletitle{Gaze-based annotations for reading comprehension}.
  In \bibinfo{booktitle}{\emph{Proceedings of the 33rd annual ACM conference on
  human factors in computing systems}}. \bibinfo{pages}{1569--1572}.
\newblock


\bibitem[Cheong et~al\mbox{.}(2007)]%
        {cheong2007relationship}
\bibfield{author}{\bibinfo{person}{Allen~MY Cheong}, \bibinfo{person}{Gordon~E
  Legge}, \bibinfo{person}{Mary~G Lawrence}, \bibinfo{person}{Sing-Hang
  Cheung}, {and} \bibinfo{person}{Mary~A Ruff}.}
  \bibinfo{year}{2007}\natexlab{}.
\newblock \showarticletitle{Relationship between slow visual processing and
  reading speed in people with macular degeneration}.
\newblock \bibinfo{journal}{\emph{Vision research}} \bibinfo{volume}{47},
  \bibinfo{number}{23} (\bibinfo{year}{2007}), \bibinfo{pages}{2943--2955}.
\newblock


\bibitem[Cheong et~al\mbox{.}(2008)]%
        {cheong2008relationship}
\bibfield{author}{\bibinfo{person}{Allen~MY Cheong}, \bibinfo{person}{Gordon~E
  Legge}, \bibinfo{person}{Mary~G Lawrence}, \bibinfo{person}{Sing-Hang
  Cheung}, {and} \bibinfo{person}{Mary~A Ruff}.}
  \bibinfo{year}{2008}\natexlab{}.
\newblock \showarticletitle{Relationship between visual span and reading
  performance in age-related macular degeneration}.
\newblock \bibinfo{journal}{\emph{Vision research}} \bibinfo{volume}{48},
  \bibinfo{number}{4} (\bibinfo{year}{2008}), \bibinfo{pages}{577--588}.
\newblock


\bibitem[Cohen(2013)]%
        {cohen2013statistical}
\bibfield{author}{\bibinfo{person}{Jacob Cohen}.}
  \bibinfo{year}{2013}\natexlab{}.
\newblock \bibinfo{booktitle}{\emph{Statistical power analysis for the
  behavioral sciences}}.
\newblock \bibinfo{publisher}{Routledge}.
\newblock


\bibitem[Crossley et~al\mbox{.}(2022)]%
        {crossley2022large}
\bibfield{author}{\bibinfo{person}{Scott Crossley}, \bibinfo{person}{Aron
  Heintz}, \bibinfo{person}{Joon~Suh Choi}, \bibinfo{person}{Jordan Batchelor},
  \bibinfo{person}{Mehrnoush Karimi}, {and} \bibinfo{person}{Agnes
  Malatinszky}.} \bibinfo{year}{2022}\natexlab{}.
\newblock \showarticletitle{A large-scaled corpus for assessing text
  readability}.
\newblock \bibinfo{journal}{\emph{Behavior Research Methods}}
  (\bibinfo{year}{2022}), \bibinfo{pages}{1--17}.
\newblock


\bibitem[Dar et~al\mbox{.}(2021)]%
        {dar2021remodnav}
\bibfield{author}{\bibinfo{person}{Asim~H Dar}, \bibinfo{person}{Adina~S
  Wagner}, {and} \bibinfo{person}{Michael Hanke}.}
  \bibinfo{year}{2021}\natexlab{}.
\newblock \showarticletitle{REMoDNaV: robust eye-movement classification for
  dynamic stimulation}.
\newblock \bibinfo{journal}{\emph{Behavior research methods}}
  \bibinfo{volume}{53}, \bibinfo{number}{1} (\bibinfo{year}{2021}),
  \bibinfo{pages}{399--414}.
\newblock


\bibitem[Durand and GOULD(1910)]%
        {durand1910method}
\bibfield{author}{\bibinfo{person}{AC Durand} {and} \bibinfo{person}{GEORGE~M
  GOULD}.} \bibinfo{year}{1910}\natexlab{}.
\newblock \showarticletitle{A method of determining ocular dominance}.
\newblock \bibinfo{journal}{\emph{Journal of the American Medical Association}}
  \bibinfo{volume}{55}, \bibinfo{number}{5} (\bibinfo{year}{1910}),
  \bibinfo{pages}{369--370}.
\newblock


\bibitem[Elkin et~al\mbox{.}(2021)]%
        {elkin2021aligned}
\bibfield{author}{\bibinfo{person}{Lisa~A Elkin}, \bibinfo{person}{Matthew
  Kay}, \bibinfo{person}{James~J Higgins}, {and} \bibinfo{person}{Jacob~O
  Wobbrock}.} \bibinfo{year}{2021}\natexlab{}.
\newblock \showarticletitle{An aligned rank transform procedure for multifactor
  contrast tests}. In \bibinfo{booktitle}{\emph{The 34th Annual ACM Symposium
  on User Interface Software and Technology}}. \bibinfo{pages}{754--768}.
\newblock


\bibitem[Engbert and Kliegl(2003)]%
        {engbert2003microsaccades}
\bibfield{author}{\bibinfo{person}{Ralf Engbert} {and}
  \bibinfo{person}{Reinhold Kliegl}.} \bibinfo{year}{2003}\natexlab{}.
\newblock \showarticletitle{Microsaccades uncover the orientation of covert
  attention}.
\newblock \bibinfo{journal}{\emph{Vision research}} \bibinfo{volume}{43},
  \bibinfo{number}{9} (\bibinfo{year}{2003}), \bibinfo{pages}{1035--1045}.
\newblock


\bibitem[Envision(2022)]%
        {envision}
\bibfield{author}{\bibinfo{person}{Envision}.} \bibinfo{year}{2022}\natexlab{}.
\newblock \bibinfo{title}{Envision App}.
\newblock
\newblock
\newblock
\shownote{Available online at: \url{https://www.letsenvision.com/app}, last
  accessed on 12/06/2022}.


\bibitem[Ferris~III et~al\mbox{.}(1982)]%
        {ferris1982new}
\bibfield{author}{\bibinfo{person}{Frederick~L Ferris~III},
  \bibinfo{person}{Aaron Kassoff}, \bibinfo{person}{George~H Bresnick}, {and}
  \bibinfo{person}{Ian Bailey}.} \bibinfo{year}{1982}\natexlab{}.
\newblock \showarticletitle{New visual acuity charts for clinical research}.
\newblock \bibinfo{journal}{\emph{American journal of ophthalmology}}
  \bibinfo{volume}{94}, \bibinfo{number}{1} (\bibinfo{year}{1982}),
  \bibinfo{pages}{91--96}.
\newblock


\bibitem[Flask(2022)]%
        {flask}
\bibfield{author}{\bibinfo{person}{Flask}.} \bibinfo{year}{2022}\natexlab{}.
\newblock \bibinfo{title}{Flask Documentation}.
\newblock
\newblock
\newblock
\shownote{Available online at:
  \url{https://flask.palletsprojects.com/en/2.2.x/}, last accessed on
  9/4/2022}.


\bibitem[Flesch(1948)]%
        {flesch1948new}
\bibfield{author}{\bibinfo{person}{Rudolph Flesch}.}
  \bibinfo{year}{1948}\natexlab{}.
\newblock \showarticletitle{A new readability yardstick.}
\newblock \bibinfo{journal}{\emph{Journal of applied psychology}}
  \bibinfo{volume}{32}, \bibinfo{number}{3} (\bibinfo{year}{1948}),
  \bibinfo{pages}{221}.
\newblock


\bibitem[Gibaldi et~al\mbox{.}(2017)]%
        {gibaldi2017evaluation}
\bibfield{author}{\bibinfo{person}{Agostino Gibaldi}, \bibinfo{person}{Mauricio
  Vanegas}, \bibinfo{person}{Peter~J Bex}, {and} \bibinfo{person}{Guido
  Maiello}.} \bibinfo{year}{2017}\natexlab{}.
\newblock \showarticletitle{Evaluation of the Tobii EyeX Eye tracking
  controller and Matlab toolkit for research}.
\newblock \bibinfo{journal}{\emph{Behavior research methods}}
  \bibinfo{volume}{49}, \bibinfo{number}{3} (\bibinfo{year}{2017}),
  \bibinfo{pages}{923--946}.
\newblock


\bibitem[Gowases et~al\mbox{.}(2011)]%
        {gowases2011text}
\bibfield{author}{\bibinfo{person}{Tersia Gowases}, \bibinfo{person}{Roman
  Bednarik}, {and} \bibinfo{person}{Markku Tukiainen}.}
  \bibinfo{year}{2011}\natexlab{}.
\newblock \showarticletitle{Text highlighting improves user experience for
  reading with magnified displays}.
\newblock In \bibinfo{booktitle}{\emph{CHI'11 Extended Abstracts on Human
  Factors in Computing Systems}}. \bibinfo{pages}{1891--1896}.
\newblock


\bibitem[Guestrin and Eizenman(2006)]%
        {guestrin2006general}
\bibfield{author}{\bibinfo{person}{Elias~Daniel Guestrin} {and}
  \bibinfo{person}{Moshe Eizenman}.} \bibinfo{year}{2006}\natexlab{}.
\newblock \showarticletitle{General theory of remote gaze estimation using the
  pupil center and corneal reflections}.
\newblock \bibinfo{journal}{\emph{IEEE Transactions on biomedical engineering}}
  \bibinfo{volume}{53}, \bibinfo{number}{6} (\bibinfo{year}{2006}),
  \bibinfo{pages}{1124--1133}.
\newblock


\bibitem[Hallett(2015)]%
        {hallett2015reading}
\bibfield{author}{\bibinfo{person}{Elyse~C Hallett}.}
  \bibinfo{year}{2015}\natexlab{}.
\newblock \bibinfo{booktitle}{\emph{Reading without bounds: How different
  magnification methods affect the performance of students with low vision}}.
\newblock \bibinfo{publisher}{California State University, Long Beach}.
\newblock


\bibitem[Hallett et~al\mbox{.}(2017)]%
        {hallett2017screen}
\bibfield{author}{\bibinfo{person}{Elyse~C Hallett}, \bibinfo{person}{Wayne
  Dick}, \bibinfo{person}{Tom Jewett}, {and} \bibinfo{person}{Kim-Phuong~L
  Vu}.} \bibinfo{year}{2017}\natexlab{}.
\newblock \showarticletitle{How screen magnification with and without
  word-wrapping affects the user experience of adults with low vision}. In
  \bibinfo{booktitle}{\emph{International Conference on Applied Human Factors
  and Ergonomics}}. Springer, \bibinfo{pages}{665--674}.
\newblock


\bibitem[Harland et~al\mbox{.}(1998)]%
        {harland1998psychophysics}
\bibfield{author}{\bibinfo{person}{Stephen Harland}, \bibinfo{person}{Gordon~E
  Legge}, {and} \bibinfo{person}{Andrew Luebker}.}
  \bibinfo{year}{1998}\natexlab{}.
\newblock \showarticletitle{Psychophysics of reading. XVII. Low-vision
  performance with four types of electronically magnified text}.
\newblock \bibinfo{journal}{\emph{Optometry and Vision Science}}
  \bibinfo{volume}{75}, \bibinfo{number}{3} (\bibinfo{year}{1998}),
  \bibinfo{pages}{183--190}.
\newblock


\bibitem[Jarodzka and Brand-Gruwel(2017)]%
        {jarodzka2017tracking}
\bibfield{author}{\bibinfo{person}{Halszka Jarodzka} {and}
  \bibinfo{person}{Saskia Brand-Gruwel}.} \bibinfo{year}{2017}\natexlab{}.
\newblock \bibinfo{title}{Tracking the reading eye: Towards a model of
  real-world reading}.
\newblock , \bibinfo{numpages}{193--201}~pages.
\newblock


\bibitem[Kar and Corcoran(2017)]%
        {kar2017review}
\bibfield{author}{\bibinfo{person}{Anuradha Kar} {and} \bibinfo{person}{Peter
  Corcoran}.} \bibinfo{year}{2017}\natexlab{}.
\newblock \showarticletitle{A review and analysis of eye-gaze estimation
  systems, algorithms and performance evaluation methods in consumer
  platforms}.
\newblock \bibinfo{journal}{\emph{IEEE Access}}  \bibinfo{volume}{5}
  (\bibinfo{year}{2017}), \bibinfo{pages}{16495--16519}.
\newblock


\bibitem[Kincaid et~al\mbox{.}(1975)]%
        {kincaid1975derivation}
\bibfield{author}{\bibinfo{person}{J~Peter Kincaid}, \bibinfo{person}{Robert~P
  Fishburne~Jr}, \bibinfo{person}{Richard~L Rogers}, {and}
  \bibinfo{person}{Brad~S Chissom}.} \bibinfo{year}{1975}\natexlab{}.
\newblock \bibinfo{booktitle}{\emph{Derivation of new readability formulas
  (automated readability index, fog count and flesch reading ease formula) for
  navy enlisted personnel}}.
\newblock \bibinfo{type}{{T}echnical {R}eport}. \bibinfo{institution}{Naval
  Technical Training Command Millington TN Research Branch}.
\newblock


\bibitem[Krischer et~al\mbox{.}(1985)]%
        {krischer1985visual}
\bibfield{author}{\bibinfo{person}{CC Krischer}, \bibinfo{person}{M
  Stein-Arsic}, \bibinfo{person}{R Meissen}, {and} \bibinfo{person}{J Zihl}.}
  \bibinfo{year}{1985}\natexlab{}.
\newblock \showarticletitle{Visual performance and reading capacity of
  partially sighted persons in a rehabilitation center.}
\newblock \bibinfo{journal}{\emph{American journal of optometry and
  physiological optics}} \bibinfo{volume}{62}, \bibinfo{number}{1}
  (\bibinfo{year}{1985}), \bibinfo{pages}{52--58}.
\newblock


\bibitem[Kwon et~al\mbox{.}(2013)]%
        {kwon2013rapid}
\bibfield{author}{\bibinfo{person}{MiYoung Kwon}, \bibinfo{person}{Anirvan~S
  Nandy}, {and} \bibinfo{person}{Bosco~S Tjan}.}
  \bibinfo{year}{2013}\natexlab{}.
\newblock \showarticletitle{Rapid and persistent adaptability of human
  oculomotor control in response to simulated central vision loss}.
\newblock \bibinfo{journal}{\emph{Current Biology}} \bibinfo{volume}{23},
  \bibinfo{number}{17} (\bibinfo{year}{2013}), \bibinfo{pages}{1663--1669}.
\newblock


\bibitem[Legge et~al\mbox{.}(1997)]%
        {legge1997psychophysics}
\bibfield{author}{\bibinfo{person}{Gordon~E Legge}, \bibinfo{person}{Sonia~J
  Ahn}, \bibinfo{person}{Timothy~S Klitz}, {and} \bibinfo{person}{Andrew
  Luebker}.} \bibinfo{year}{1997}\natexlab{}.
\newblock \showarticletitle{Psychophysics of reading—XVI. The visual span in
  normal and low vision}.
\newblock \bibinfo{journal}{\emph{Vision Research}} \bibinfo{volume}{37},
  \bibinfo{number}{14} (\bibinfo{year}{1997}), \bibinfo{pages}{1999--2010}.
\newblock


\bibitem[Legge et~al\mbox{.}(1985)]%
        {legge1985psychophysics}
\bibfield{author}{\bibinfo{person}{Gordon~E Legge}, \bibinfo{person}{Gary~S
  Rubin}, \bibinfo{person}{Denis~G Pelli}, {and} \bibinfo{person}{Mary~M
  Schleske}.} \bibinfo{year}{1985}\natexlab{}.
\newblock \showarticletitle{Psychophysics of reading—II. Low vision}.
\newblock \bibinfo{journal}{\emph{Vision research}} \bibinfo{volume}{25},
  \bibinfo{number}{2} (\bibinfo{year}{1985}), \bibinfo{pages}{253--265}.
\newblock


\bibitem[Ltd.(2012)]%
        {eyelink}
\bibfield{author}{\bibinfo{person}{SR~Research Ltd.}}
  \bibinfo{year}{2012}\natexlab{}.
\newblock \bibinfo{title}{EyeLink 1000 Quotation}.
\newblock
\newblock
\newblock
\shownote{Available online at:
  \url{https://techfee.fau.edu/approvedproposals/Download.cfm?sid=7&pid=44},
  last accessed on 9/15/2022}.


\bibitem[Masnadi et~al\mbox{.}(2020)]%
        {masnadi2020vriassist}
\bibfield{author}{\bibinfo{person}{Sina Masnadi}, \bibinfo{person}{Brian
  Williamson}, \bibinfo{person}{Andr{\'e}s N~Vargas Gonz{\'a}lez}, {and}
  \bibinfo{person}{Joseph~J LaViola}.} \bibinfo{year}{2020}\natexlab{}.
\newblock \showarticletitle{Vriassist: An eye-tracked virtual reality low
  vision assistance tool}. In \bibinfo{booktitle}{\emph{2020 IEEE Conference on
  Virtual Reality and 3D User Interfaces Abstracts and Workshops (VRW)}}. IEEE,
  \bibinfo{pages}{808--809}.
\newblock


\bibitem[Maus et~al\mbox{.}(2020)]%
        {maus2020gaze}
\bibfield{author}{\bibinfo{person}{Natalie Maus}, \bibinfo{person}{Dalton
  Rutledge}, \bibinfo{person}{Sedeeq Al-Khazraji}, \bibinfo{person}{Reynold
  Bailey}, \bibinfo{person}{Cecilia~Ovesdotter Alm}, {and}
  \bibinfo{person}{Kristen Shinohara}.} \bibinfo{year}{2020}\natexlab{}.
\newblock \showarticletitle{Gaze-guided Magnification for Individuals with
  Vision Impairments}. In \bibinfo{booktitle}{\emph{Extended Abstracts of the
  2020 CHI Conference on Human Factors in Computing Systems}}.
  \bibinfo{pages}{1--8}.
\newblock


\bibitem[Mc~Laughlin(1969)]%
        {mc1969smog}
\bibfield{author}{\bibinfo{person}{G~Harry Mc~Laughlin}.}
  \bibinfo{year}{1969}\natexlab{}.
\newblock \showarticletitle{SMOG grading-a new readability formula}.
\newblock \bibinfo{journal}{\emph{Journal of reading}} \bibinfo{volume}{12},
  \bibinfo{number}{8} (\bibinfo{year}{1969}), \bibinfo{pages}{639--646}.
\newblock


\bibitem[Microsoft(2022)]%
        {winmag}
\bibfield{author}{\bibinfo{person}{Microsoft}.}
  \bibinfo{year}{2022}\natexlab{}.
\newblock \bibinfo{title}{Use Magnifier to make things on the screen easier to
  see}.
\newblock
\newblock
\newblock
\shownote{Available online at:
  \url{https://support.microsoft.com/en-us/windows/use-magnifier-to-make-things-on-the-screen-easier-to-see-414948ba-8b1c-d3bd-8615-0e5e32204198\#WindowsVersion=Windows_11},
  last accessed on 8/23/2022}.


\bibitem[Moffitt(1980)]%
        {moffitt1980evaluation}
\bibfield{author}{\bibinfo{person}{Kirk Moffitt}.}
  \bibinfo{year}{1980}\natexlab{}.
\newblock \showarticletitle{Evaluation of the fixation duration in visual
  search}.
\newblock \bibinfo{journal}{\emph{Perception \& Psychophysics}}
  \bibinfo{volume}{27}, \bibinfo{number}{4} (\bibinfo{year}{1980}),
  \bibinfo{pages}{370--372}.
\newblock


\bibitem[Moreno et~al\mbox{.}(2021)]%
        {moreno2021exploratory}
\bibfield{author}{\bibinfo{person}{Lourdes Moreno}, \bibinfo{person}{Xabier
  Valencia}, \bibinfo{person}{J~Eduardo P{\'e}rez}, {and}
  \bibinfo{person}{Myriam Arrue}.} \bibinfo{year}{2021}\natexlab{}.
\newblock \showarticletitle{An exploratory study of web adaptation techniques
  for people with low vision}.
\newblock \bibinfo{journal}{\emph{Universal access in the information society}}
  \bibinfo{volume}{20}, \bibinfo{number}{2} (\bibinfo{year}{2021}),
  \bibinfo{pages}{223--237}.
\newblock


\bibitem[Murai et~al\mbox{.}(2010)]%
        {murai2010eye}
\bibfield{author}{\bibinfo{person}{Yasuyuki Murai}, \bibinfo{person}{Masaji
  Kawahara}, \bibinfo{person}{Hisayuki Tatsumi}, \bibinfo{person}{Iwao Sekita},
  {and} \bibinfo{person}{Masahiro Miyakawa}.} \bibinfo{year}{2010}\natexlab{}.
\newblock \showarticletitle{Eye tracking for low vision aids-toward guiding of
  gaze}. In \bibinfo{booktitle}{\emph{International conference on computers for
  handicapped persons}}. Springer, \bibinfo{pages}{308--315}.
\newblock


\bibitem[NIH(2020)]%
        {nei}
\bibfield{author}{\bibinfo{person}{NIH}.} \bibinfo{year}{2020}\natexlab{}.
\newblock \bibinfo{title}{Low Vision - National Eye Institute}.
\newblock
\newblock
\newblock
\shownote{Available online at:
  \url{https://www.nei.nih.gov/learn-about-eye-health/eye-conditions-and-diseases/low-vision},
  last accessed on 8/23/2022}.


\bibitem[Pagon(1981)]%
        {pagon1981ocular}
\bibfield{author}{\bibinfo{person}{Roberta~A Pagon}.}
  \bibinfo{year}{1981}\natexlab{}.
\newblock \showarticletitle{Ocular coloboma}.
\newblock \bibinfo{journal}{\emph{Survey of ophthalmology}}
  \bibinfo{volume}{25}, \bibinfo{number}{4} (\bibinfo{year}{1981}),
  \bibinfo{pages}{223--236}.
\newblock


\bibitem[Prahalad and Coates(2020)]%
        {prahalad2020asymmetries}
\bibfield{author}{\bibinfo{person}{Krishnamachari~S Prahalad} {and}
  \bibinfo{person}{Daniel~R Coates}.} \bibinfo{year}{2020}\natexlab{}.
\newblock \showarticletitle{Asymmetries of reading eye movements in simulated
  central vision loss}.
\newblock \bibinfo{journal}{\emph{Vision research}}  \bibinfo{volume}{171}
  (\bibinfo{year}{2020}), \bibinfo{pages}{1--10}.
\newblock


\bibitem[Pro(2021)]%
        {tobiiproacc}
\bibfield{author}{\bibinfo{person}{Tobii Pro}.}
  \bibinfo{year}{2021}\natexlab{}.
\newblock \bibinfo{title}{Tobii Pro Fusion Field Metrics Test Report}.
\newblock
\newblock
\newblock
\shownote{Available online at:
  \url{https://www.tobiipro.com/siteassets/tobii-pro/products/hardware/fusion/tobii-pro-fusion-field-metrics-test-report.pdf},
  last accessed on 9/4/2022}.


\bibitem[Pro(2022a)]%
        {tobiicali}
\bibfield{author}{\bibinfo{person}{Tobii Pro}.}
  \bibinfo{year}{2022}\natexlab{a}.
\newblock \bibinfo{title}{Performing a calibration and validation in Pro Lab}.
\newblock
\newblock
\newblock
\shownote{Available online at:
  \url{https://www.tobiipro.com/learn-and-support/learn/steps-in-an-eye-tracking-study/run/performing-a-calibration-and-validation-in-pro-lab/},
  last accessed on 9/7/2022}.


\bibitem[Pro(2022b)]%
        {tobiiprofusion}
\bibfield{author}{\bibinfo{person}{Tobii Pro}.}
  \bibinfo{year}{2022}\natexlab{b}.
\newblock \bibinfo{title}{Screen-based eye tracker for research | Tobii Pro
  Fusion}.
\newblock
\newblock
\newblock
\shownote{Available online at:
  \url{https://www.tobiipro.com/product-listing/fusion/}, last accessed on
  9/7/2022}.


\bibitem[Pro(2022c)]%
        {tobiiprolab}
\bibfield{author}{\bibinfo{person}{Tobii Pro}.}
  \bibinfo{year}{2022}\natexlab{c}.
\newblock \bibinfo{title}{Tobii Pro Lab}.
\newblock
\newblock
\newblock
\shownote{Available online at:
  \url{https://www.tobiipro.com/product-listing/tobii-pro-lab/}, last accessed
  on 9/7/2022}.


\bibitem[Pro(2022d)]%
        {tobiiprosdk}
\bibfield{author}{\bibinfo{person}{Tobii Pro}.}
  \bibinfo{year}{2022}\natexlab{d}.
\newblock \bibinfo{title}{Tobii Pro SDK}.
\newblock
\newblock
\newblock
\shownote{Available online at:
  \url{https://www.tobiipro.com/product-listing/tobii-pro-sdk/}, last accessed
  on 9/7/2022}.


\bibitem[Racette et~al\mbox{.}(2016)]%
        {racette2016visual}
\bibfield{author}{\bibinfo{person}{L Racette}, \bibinfo{person}{M Fischer},
  \bibinfo{person}{H Bebie}, \bibinfo{person}{G Holl{\'o}}, \bibinfo{person}{CA
  Johnson}, {and} \bibinfo{person}{C Matsumoto}.}
  \bibinfo{year}{2016}\natexlab{}.
\newblock \showarticletitle{Visual field digest: A guide to perimetry and the
  Octopus perimeter}.
\newblock \bibinfo{journal}{\emph{K{\"o}niz, Switzerland: Haag-Streit AG}}
  \bibinfo{volume}{289} (\bibinfo{year}{2016}).
\newblock


\bibitem[Rayner(1998)]%
        {rayner1998eye}
\bibfield{author}{\bibinfo{person}{Keith Rayner}.}
  \bibinfo{year}{1998}\natexlab{}.
\newblock \showarticletitle{Eye movements in reading and information
  processing: 20 years of research.}
\newblock \bibinfo{journal}{\emph{Psychological bulletin}}
  \bibinfo{volume}{124}, \bibinfo{number}{3} (\bibinfo{year}{1998}),
  \bibinfo{pages}{372}.
\newblock


\bibitem[Rayner(2009)]%
        {rayner200935th}
\bibfield{author}{\bibinfo{person}{Keith Rayner}.}
  \bibinfo{year}{2009}\natexlab{}.
\newblock \showarticletitle{The 35th Sir Frederick Bartlett Lecture: Eye
  movements and attention in reading, scene perception, and visual search}.
\newblock \bibinfo{journal}{\emph{Quarterly journal of experimental
  psychology}} \bibinfo{volume}{62}, \bibinfo{number}{8}
  (\bibinfo{year}{2009}), \bibinfo{pages}{1457--1506}.
\newblock


\bibitem[Rayner et~al\mbox{.}(2012)]%
        {rayner2012psychology}
\bibfield{author}{\bibinfo{person}{Keith Rayner}, \bibinfo{person}{Alexander
  Pollatsek}, \bibinfo{person}{Jane Ashby}, {and} \bibinfo{person}{Charles
  Clifton~Jr}.} \bibinfo{year}{2012}\natexlab{}.
\newblock \bibinfo{booktitle}{\emph{Psychology of reading}}.
\newblock \bibinfo{publisher}{Psychology Press}.
\newblock


\bibitem[Rayner et~al\mbox{.}(2010)]%
        {rayner2010eye}
\bibfield{author}{\bibinfo{person}{Keith Rayner}, \bibinfo{person}{Timothy~J
  Slattery}, {and} \bibinfo{person}{Nathalie~N B{\'e}langer}.}
  \bibinfo{year}{2010}\natexlab{}.
\newblock \showarticletitle{Eye movements, the perceptual span, and reading
  speed}.
\newblock \bibinfo{journal}{\emph{Psychonomic bulletin \& review}}
  \bibinfo{volume}{17}, \bibinfo{number}{6} (\bibinfo{year}{2010}),
  \bibinfo{pages}{834--839}.
\newblock


\bibitem[Reichle et~al\mbox{.}(2003)]%
        {reichle2003ez}
\bibfield{author}{\bibinfo{person}{Erik~D Reichle}, \bibinfo{person}{Keith
  Rayner}, {and} \bibinfo{person}{Alexander Pollatsek}.}
  \bibinfo{year}{2003}\natexlab{}.
\newblock \showarticletitle{The EZ Reader model of eye-movement control in
  reading: Comparisons to other models}.
\newblock \bibinfo{journal}{\emph{Behavioral and brain sciences}}
  \bibinfo{volume}{26}, \bibinfo{number}{4} (\bibinfo{year}{2003}),
  \bibinfo{pages}{445--476}.
\newblock


\bibitem[Renninger and Ma-Wyatt(2011)]%
        {renninger2011recalibration}
\bibfield{author}{\bibinfo{person}{Laura Renninger} {and} \bibinfo{person}{Anna
  Ma-Wyatt}.} \bibinfo{year}{2011}\natexlab{}.
\newblock \showarticletitle{Recalibration of eye and hand reference frames in
  age-related macular degeneration}.
\newblock \bibinfo{journal}{\emph{Journal of Vision}} \bibinfo{volume}{11},
  \bibinfo{number}{11} (\bibinfo{year}{2011}), \bibinfo{pages}{954--954}.
\newblock


\bibitem[Robinson(1963)]%
        {robinson1963method}
\bibfield{author}{\bibinfo{person}{David~A Robinson}.}
  \bibinfo{year}{1963}\natexlab{}.
\newblock \showarticletitle{A method of measuring eye movemnent using a scieral
  search coil in a magnetic field}.
\newblock \bibinfo{journal}{\emph{IEEE Transactions on bio-medical
  electronics}} \bibinfo{volume}{10}, \bibinfo{number}{4}
  (\bibinfo{year}{1963}), \bibinfo{pages}{137--145}.
\newblock


\bibitem[Rubin and Feely(2009)]%
        {rubin2009role}
\bibfield{author}{\bibinfo{person}{Gary~S Rubin} {and} \bibinfo{person}{Mary
  Feely}.} \bibinfo{year}{2009}\natexlab{}.
\newblock \showarticletitle{The role of eye movements during reading in
  patients with age-related macular degeneration (AMD)}.
\newblock \bibinfo{journal}{\emph{Neuro-Ophthalmology}} \bibinfo{volume}{33},
  \bibinfo{number}{3} (\bibinfo{year}{2009}), \bibinfo{pages}{120--126}.
\newblock


\bibitem[Salda{\~n}a(2021)]%
        {saldana2021coding}
\bibfield{author}{\bibinfo{person}{Johnny Salda{\~n}a}.}
  \bibinfo{year}{2021}\natexlab{}.
\newblock \showarticletitle{The coding manual for qualitative researchers}.
\newblock \bibinfo{journal}{\emph{The coding manual for qualitative
  researchers}} (\bibinfo{year}{2021}), \bibinfo{pages}{1--440}.
\newblock


\bibitem[Salvucci and Goldberg(2000)]%
        {salvucci2000identifying}
\bibfield{author}{\bibinfo{person}{Dario~D Salvucci} {and}
  \bibinfo{person}{Joseph~H Goldberg}.} \bibinfo{year}{2000}\natexlab{}.
\newblock \showarticletitle{Identifying fixations and saccades in eye-tracking
  protocols}. In \bibinfo{booktitle}{\emph{Proceedings of the 2000 symposium on
  Eye tracking research \& applications}}. \bibinfo{pages}{71--78}.
\newblock


\bibitem[Shanidze and Velisar(2020)]%
        {shanidze2020eye}
\bibfield{author}{\bibinfo{person}{Natela~M Shanidze} {and}
  \bibinfo{person}{Anca Velisar}.} \bibinfo{year}{2020}\natexlab{}.
\newblock \showarticletitle{Eye, head, and gaze contributions to smooth pursuit
  in macular degeneration}.
\newblock \bibinfo{journal}{\emph{Journal of Neurophysiology}}
  \bibinfo{volume}{124}, \bibinfo{number}{1} (\bibinfo{year}{2020}),
  \bibinfo{pages}{134--144}.
\newblock


\bibitem[Simonsz and Bour(1991)]%
        {simonsz1991covering}
\bibfield{author}{\bibinfo{person}{HJ Simonsz} {and} \bibinfo{person}{LJ
  Bour}.} \bibinfo{year}{1991}\natexlab{}.
\newblock \showarticletitle{Covering one eye in fixation-disparity measurement
  causes slight movement of fellow eye}.
\newblock \bibinfo{journal}{\emph{Documenta Ophthalmologica}}
  \bibinfo{volume}{78}, \bibinfo{number}{3} (\bibinfo{year}{1991}),
  \bibinfo{pages}{141--152}.
\newblock


\bibitem[Source(2022)]%
        {react}
\bibfield{author}{\bibinfo{person}{Meta~Open Source}.}
  \bibinfo{year}{2022}\natexlab{}.
\newblock \bibinfo{title}{React - A JavaScript library for building user
  interfaces}.
\newblock
\newblock
\newblock
\shownote{Available online at: \url{https://reactjs.org}, last accessed on
  9/7/2022}.


\bibitem[Szpiro et~al\mbox{.}(2016)]%
        {szpiro2016people}
\bibfield{author}{\bibinfo{person}{Sarit Felicia~Anais Szpiro},
  \bibinfo{person}{Shafeka Hashash}, \bibinfo{person}{Yuhang Zhao}, {and}
  \bibinfo{person}{Shiri Azenkot}.} \bibinfo{year}{2016}\natexlab{}.
\newblock \showarticletitle{How people with low vision access computing
  devices: Understanding challenges and opportunities}. In
  \bibinfo{booktitle}{\emph{Proceedings of the 18th International ACM SIGACCESS
  Conference on Computers and Accessibility}}. \bibinfo{pages}{171--180}.
\newblock


\bibitem[Verghese et~al\mbox{.}(2021)]%
        {verghese2021eye}
\bibfield{author}{\bibinfo{person}{Preeti Verghese},
  \bibinfo{person}{C{\'e}cile Vullings}, {and} \bibinfo{person}{Natela
  Shanidze}.} \bibinfo{year}{2021}\natexlab{}.
\newblock \showarticletitle{Eye Movements in Macular Degeneration}.
\newblock \bibinfo{journal}{\emph{Annual Review of Vision Science}}
  \bibinfo{volume}{7} (\bibinfo{year}{2021}), \bibinfo{pages}{773--791}.
\newblock


\bibitem[Vickers(2009)]%
        {vickers2009advances}
\bibfield{author}{\bibinfo{person}{Joan~N Vickers}.}
  \bibinfo{year}{2009}\natexlab{}.
\newblock \showarticletitle{Advances in coupling perception and action: the
  quiet eye as a bidirectional link between gaze, attention, and action}.
\newblock \bibinfo{journal}{\emph{Progress in brain research}}
  \bibinfo{volume}{174} (\bibinfo{year}{2009}), \bibinfo{pages}{279--288}.
\newblock


\bibitem[W3C(2022)]%
        {wcag}
\bibfield{author}{\bibinfo{person}{W3C}.} \bibinfo{year}{2022}\natexlab{}.
\newblock \bibinfo{title}{Web Content Accessibility Guidelines (WCAG) 2.1}.
\newblock
\newblock
\newblock
\shownote{Available online at:
  \url{https://www.w3.org/TR/WCAG21/\#text-spacing}, last accessed on
  9/7/2022}.


\bibitem[Wobbrock et~al\mbox{.}(2011)]%
        {wobbrock2011aligned}
\bibfield{author}{\bibinfo{person}{Jacob~O Wobbrock}, \bibinfo{person}{Leah
  Findlater}, \bibinfo{person}{Darren Gergle}, {and} \bibinfo{person}{James~J
  Higgins}.} \bibinfo{year}{2011}\natexlab{}.
\newblock \showarticletitle{The aligned rank transform for nonparametric
  factorial analyses using only anova procedures}. In
  \bibinfo{booktitle}{\emph{Proceedings of the SIGCHI conference on human
  factors in computing systems}}. \bibinfo{pages}{143--146}.
\newblock


\bibitem[Zambarbieri and Carniglia(2012)]%
        {zambarbieri2012eye}
\bibfield{author}{\bibinfo{person}{Daniela Zambarbieri} {and}
  \bibinfo{person}{Elena Carniglia}.} \bibinfo{year}{2012}\natexlab{}.
\newblock \showarticletitle{Eye movement analysis of reading from computer
  displays, eReaders and printed books}.
\newblock \bibinfo{journal}{\emph{Ophthalmic and Physiological Optics}}
  \bibinfo{volume}{32}, \bibinfo{number}{5} (\bibinfo{year}{2012}),
  \bibinfo{pages}{390--396}.
\newblock


\bibitem[Zhang et~al\mbox{.}(2021)]%
        {zhang2021eye}
\bibfield{author}{\bibinfo{person}{Xucong Zhang}, \bibinfo{person}{Seonwook
  Park}, {and} \bibinfo{person}{Anna Maria~Feit}.}
  \bibinfo{year}{2021}\natexlab{}.
\newblock \showarticletitle{Eye Gaze Estimation and Its Applications}.
\newblock In \bibinfo{booktitle}{\emph{Artificial Intelligence for Human
  Computer Interaction: A Modern Approach}}. \bibinfo{publisher}{Springer},
  \bibinfo{pages}{99--130}.
\newblock


\bibitem[Zhao et~al\mbox{.}(2017)]%
        {zhao2017understanding}
\bibfield{author}{\bibinfo{person}{Yuhang Zhao}, \bibinfo{person}{Michele Hu},
  \bibinfo{person}{Shafeka Hashash}, {and} \bibinfo{person}{Shiri Azenkot}.}
  \bibinfo{year}{2017}\natexlab{}.
\newblock \showarticletitle{Understanding low vision people's visual perception
  on commercial augmented reality glasses}. In
  \bibinfo{booktitle}{\emph{Proceedings of the 2017 CHI Conference on Human
  Factors in Computing Systems}}. \bibinfo{pages}{4170--4181}.
\newblock


\bibitem[Zhao et~al\mbox{.}(2015)]%
        {zhao2015foresee}
\bibfield{author}{\bibinfo{person}{Yuhang Zhao}, \bibinfo{person}{Sarit
  Szpiro}, {and} \bibinfo{person}{Shiri Azenkot}.}
  \bibinfo{year}{2015}\natexlab{}.
\newblock \showarticletitle{Foresee: A customizable head-mounted vision
  enhancement system for people with low vision}. In
  \bibinfo{booktitle}{\emph{Proceedings of the 17th international ACM SIGACCESS
  conference on computers \& accessibility}}. \bibinfo{pages}{239--249}.
\newblock


\bibitem[Zhu and Ji(2007)]%
        {zhu2007novel}
\bibfield{author}{\bibinfo{person}{Zhiwei Zhu} {and} \bibinfo{person}{Qiang
  Ji}.} \bibinfo{year}{2007}\natexlab{}.
\newblock \showarticletitle{Novel eye gaze tracking techniques under natural
  head movement}.
\newblock \bibinfo{journal}{\emph{IEEE Transactions on biomedical engineering}}
  \bibinfo{volume}{54}, \bibinfo{number}{12} (\bibinfo{year}{2007}),
  \bibinfo{pages}{2246--2260}.
\newblock


\end{thebibliography}

\end{document}